\documentclass[12pt]{article}

\usepackage{cite}
\usepackage{amsmath}
\usepackage{float} 
\usepackage{graphicx}
\usepackage{amsfonts}
\usepackage{longtable}
\usepackage{tikz}

\usetikzlibrary{arrows.meta, decorations.markings, positioning}
\input{epsf}

\usepackage[colorlinks,linkcolor=blue,citecolor=red,hyperindex]{hyperref}

\def\rf#1{(\ref{eq:#1})}
\def\lab#1{\label{eq:#1}}

\def\br{\begin{eqnarray}}
\def\er{\end{eqnarray}}
\def\be{\begin{equation}}
\def\ee{\end{equation}}

\def\({\left(}
\def\){\right)}

\def\rlx{\relax\leavevmode}
\def\IR{\rlx\hbox{\rm I\kern-.18em R}}
\def\vp{\varphi}

\def\u2{\mid u\mid^2}

\relax

\def\Tr{\mathop{\rm Tr}}

\newcommand{\sbr}[2]{\left\lbrack\,{#1}\, ,\,{#2}\,\right\rbrack}
\newcommand{\pbr}[2]{\{\,{#1}\, ,\,{#2}\,\}}
\newcommand{\pbro}[2]{\{\,{#1}\,\overset{\otimes}{,}\,{#2}\,\}}

\def\a{\alpha}

\def\b{\beta}

\def\d{\delta}

\def\vareps{\varepsilon}
\def\g{\gamma}
\def\G{\Gamma}

\def\l{\lambda}

\def\L{\Lambda}
\def\m{\mu}

\def\n{\nu}

\def\s{\sigma}
\def\S{\Sigma}
\def\t{\tau}
\def\th{\theta}

\def\tp0{\Theta_{+}^{(0)}}
\def\tm0{\Theta_{-}^{(0)}}

\def\u2{\mid u\mid^2}

\def\wti{\widetilde}

\def\vp{\varphi}

\def\rlx{\relax\leavevmode}
\def\inbar{\vrule height1.5ex width.4pt depth0pt}
\def\IZ{\rlx\hbox{\sf Z\kern-.4em Z}}
\def\IR{\rlx\hbox{\rm I\kern-.18em R}}
\def\IC{\rlx\hbox{\,$\inbar\kern-.3em{\rm C}$}}
\def\IN{\rlx\hbox{\rm I\kern-.18em N}}
\def\IO{\rlx\hbox{\,$\inbar\kern-.3em{\rm O}$}}
\def\IP{\rlx\hbox{\rm I\kern-.18em P}}
\def\IQ{\rlx\hbox{\,$\inbar\kern-.3em{\rm Q}$}}
\def\IF{\rlx\hbox{\rm I\kern-.18em F}}
\def\IG{\rlx\hbox{\,$\inbar\kern-.3em{\rm G}$}}
\def\IH{\rlx\hbox{\rm I\kern-.18em H}}
\def\II{\rlx\hbox{\rm I\kern-.18em I}}
\def\IK{\rlx\hbox{\rm I\kern-.18em K}}
\def\IL{\rlx\hbox{\rm I\kern-.18em L}}
\def\one{\hbox{{1}\kern-.25em\hbox{l}}}
\def\0#1{\relax\ifmmode\mathaccent''7017{#1}%
B        \else\accent23#1\relax\fi}

\begin{document}

\begin{titlepage}
\vspace*{-1cm}

\vskip 2cm

\vspace{.2in}
\begin{center}
{\large\bf The Hidden Symmetries of Yang-Mills Theory in $(1+1)$-dimensions}
\end{center}

\begin{center}

L. A. Ferreira$^{\dagger ,}$\footnote{laf@ifsc.usp.br}, 
G. Luchini$^{\star ,}$\footnote{gabriel.luchini@ufes.br}, H. Malavazzi$^{\dagger ,}$\footnote{henrique.malavazzi@usp.br}

\small

\par \vskip .2in \noindent

$^{\dagger}$Instituto de F\'\i sica de S\~ao Carlos; IFSC/USP;\\
Universidade de S\~ao Paulo, USP  \\ 
Caixa Postal 369, CEP 13560-970, S\~ao Carlos-SP, Brazil\\

\par \vskip .2in \noindent 
$^{\star}$ Departamento de F\'isica\\
 Universidade Federal do Esp\'irito Santo (UFES),\\
 CEP 29075-900, Vit\'oria-ES, Brazil

\vskip 1cm

\begin{abstract}
We present an integral formulation of classical Yang-Mills theory coupled to fermionic and scalar matter fields in $(1+1)$-dimensional Minkowski spacetime. By reformulating the local dynamics in terms of loop-space holonomies, we demonstrate that the path independence of the holonomy eigenvalues constitutes a conservation law, yielding an infinite hierarchy of gauge-invariant, dynamically conserved charges. While a zero-curvature equation is associated with a necessary condition for this path invariance, we note that it is not strictly sufficient on its own. Employing a first-order symplectic formalism, we show that these non-abelian charges generate global symmetry transformations on the fundamental phase-space variables. We rigorously prove that these transformations preserve the physical dynamics, leaving the total Hamiltonian invariant up to first-class constraints. Furthermore, an analysis of the Poisson algebra reveals that these conserved charges are in involution, provided the boundary integration constant lies within the center of the gauge group. This exact, lower-dimensional framework provides a highly tractable setting to investigate the algebraic structures of these hidden symmetries and the meaning of the conserved charges as physical observables, establishing a classical foundation for exploring their role in the quantum regime, such as in strongly coupled lattice gauge theories.

\end{abstract}

\normalsize
\end{center}
\end{titlepage}
\section{Introduction}
\label{sec:introduction}
\setcounter{equation}{0}
The construction of gauge-invariant conserved charges is a central and subtle problem in Yang-Mills theory \cite{ymoriginal, POLYAKOV1980171,Dolan:1980qw}. One fruitful approach to it is to view charges as being correlated with fluxes of fields. This idea lies at the very foundation of classical electrodynamics as Maxwell's theory emerged not from differential field equations but from a set of integral relations for the flux of the electromagnetic field through closed space-time surfaces.  These integral equations not only suggested a geometric interpretation of the physical quantities but also encoded a robust conservation law for the electric charge. 

Motivated by this viewpoint, the definition of charges in Yang-Mills theories requires the formulation of integral equations for field fluxes that remain valid for any choice of gauge. This can be achieved by moving beyond local field variables and adopting a framework in which the gauge field evaluated along loops, rather than at points, becomes the fundamental object so that the gauge principle is naturally reformulated in terms of holonomies of the Yang–Mills connection \cite{lafgl1, lafgl2}. 

Within this approach, the Yang–Mills equations admit an integral formulation derived from a non-abelian Stokes theorem for holonomies. A key consequence of this formulation is a path independence condition, which can be viewed as a zero-curvature condition for a connection defined over the so-called loop space - the space of functions that map closed submanifolds into space-time \cite{afg1, afg2}. This condition plays the role of a conservation law and leads to the construction of gauge-invariant conserved charges \cite{lafgl1, lafgl2}.

The path independence condition leads to an infinite set of non-abelian electric and magnetic charges, as well as their higher-order modes \cite{lafgl1,lafgl2}. In particular, recent results \cite{ymintg} have shown that the charges derived from the integral formulation of the Yang-Mills equations are in involution, thus establishing an underlying integrable structure in these theories. Moreover, these charges generate global transformations that leave the Hamiltonian invariant, thus revealing a previously unknown hidden symmetry of the physical system.

The role of the non-abelian conserved charges and their associated symmetries remains unclear in the context of the Standard Model. Although these quantities emerge naturally in the classical formulation of gauge theories, their relevance at the quantum level, particularly in connection with non-perturbative methods, is yet to be understood. The ($1+1$)-dimensional Yang-Mills theory provides an ideal setting for investigating the structure and implications of these charges in a controlled and exactly solvable context \cite{twolattice}. 

In this work, we extend the approach developed in \cite{lafgl1, lafgl2, ymintg} to the case of classical Yang-Mills theories in two-dimensional space-time coupled to spin-$\tfrac{1}{2}$ fermions and scalar bosons. We derive the Yang-Mills equations in integral form, identify their associated conserved gauge-invariant quantities, and investigate the transformations of the physical fields generated by these charges. 
We find that the transformations on the physical fields generated by the charges correspond to non-local phase factors interpreted as the parallel transport of a global charge operator by a Wilson line to the space-time point where the fields are evaluated.

Although significant progress has been made in recent years in understanding QCD in $(3+1)$ dimensions, computing the correlators of classical conserved charges in four-dimensional Yang-Mills theory remains a technically difficult problem. To address this challenge, we focus on constructing analogous structures in the more tractable setting of the two-dimensional classical Yang-Mills theory. In particular, the classical structures developed here may provide a useful starting point for investigating their quantum counterparts. For example, studying the correlators of these charges in the two-dimensional lattice QCD \cite{twolattice} can offer valuable information, especially given the solvability of the quantum regime in two dimensions \cite{Ashtekar:1996nn}.

The paper is organized as follows: In Section \ref{sec:2}, we present the integral formulation of the dynamical equations and construct the dynamically conserved, gauge-invariant charges for the $(1+1)$-dimensional Yang-Mills theory coupled to color-charged bosonic and fermionic fields. In Sections \ref{sec:3} and \ref{sec:4}, we introduce the first-order Hamiltonian formalism for the two-dimensional theory and define the charge operator that yields an infinite hierarchy of conserved charges. We then investigate the symmetries generated by these charges through the phase space Poisson structure. In Section \ref{sec:5}, we compute the Poisson algebra of the conserved charges, establishing the conditions under which they commute and are thus in involution. In Section \ref{sec:6}, we also demonstrate the existence of a hidden symmetry that preserves the conservation law, guaranteeing the existence of charges associated with a zero-curvature representation. The symmetry has a group structure that may play a role analogous to that of Kac–Moody groups in integrable models.  Finally, in Section \ref{sec:7}, we present our concluding remarks and discuss the physical consequences of our results.

\section{The integral Yang-Mills equations in two-dimensional spacetime}
\label{sec:2}
\setcounter{equation}{0}

In four-dimensional Minkowski space-time $\mathcal{M}$, with coordinates  $x^\mu$ ( $\mu = 0,1,2,3$) and metric tensor $\eta_{\mu\nu} = \textrm{diag}(1,-1,-1,-1)$, the Yang-Mills gauge field is a Lorentz vector $A_{\mu} = A_{\mu}^a\,T_a$ taking values in the Lie algebra of a compact gauge group $G$ of dimension $N$, with generators $T_a$ ($a=1,2,\dots,N$) satisfying the commutation relations 
\be
\sbr{T_a}{T_b} = i\, f_{abc}\,T_c, \qquad {\rm Tr}\(T_a\,T_b\) = \delta_{ab}.
\ee
We will consider the action of the system as follows
\be
S = \int d^2x \bigg\{-\frac{1}{4}\,{\rm Tr}\(F_{\m \n}\,F^{\m \n}\)  +{\bar \psi}^{(f)}\(i\,\gamma^{\mu}\,D_{\mu}-M_{(f)}\)\psi^{(f)} + (D_\mu \varphi)^\dagger D_{\m}\varphi - V(|\varphi|)
\bigg\}
\ee
where the covariant derivatives are given by
\br
D_\mu \psi^{(f)} &=& \partial_\mu \psi^{(f)} + i\,e\,R_{\psi}(A_\mu)\,\psi^{(f)},\nonumber\\
D_\mu \vp &=& \partial_\mu \vp + i\,e\,R_{\phi}(A_\mu)\,\vp.
\er
with $e$ the coupling constant.

The matter fields include spinor fields $\psi^{(f)}$ (fermion) with flavor index $f$, and scalar fields $\varphi$ (boson), with $R_\psi$, $R_\phi$ denoting the representations of $G$ under which the fermions and bosons transform, respectively.

So, the dynamics of the gauge field is governed by the Yang-Mills equations,
\be
D_\mu F^{\mu \nu} = e\,J^\nu, \qquad 
D_\mu {\widetilde F}^{\mu \nu} = 0,
\lab{sec2:deqym}
\ee
where the field strength tensor $F_{\mu \nu}$ is defined by 
\be
F_{\mu \nu} = \partial_\mu A_\nu - \partial_\nu A_\mu + i\,e\,\sbr{A_\mu}{A_\nu}
\ee
and ${\widetilde F}_{\mu \nu} =\frac{1}{2} \vareps_{\mu \nu \sigma \lambda} F^{\sigma \lambda}$ stands for its Hodge-dual\footnote{We use $\vareps_{0123} = 1$.}.
The interaction with charged-matter fields is described by the current $J^\mu$, which serves as the source term in the Yang-Mills equations, given by 
\be
J_\mu = J^\psi_\mu + J^\vp_\mu, \qquad \qquad \qquad \mu=0,1,2,3
\lab{curr}
\ee
with 
\br
J^\psi_\mu &\equiv& \bar{\psi}\,\gamma_\mu\,R_{\psi}(T_a)\,\psi\,T_a,\nonumber\\
J^\vp_\mu &\equiv& -i\,[\(D_\mu \vp\)^\dagger\(R_\vp(T_a)\,\vp\) - \(\vp^\dagger\,R_{\vp}(T_a)\) D_\mu \vp ]\,T_a.
\er

The Yang-Mills equations in the $1+1$ dimensional spacetime can be obtained by performing a dimensional reduction of \rf{sec2:deqym}, setting the transverse components of the gauge field $A^i$, $i=2,3$ to zero and restricting the remaining components to depend only on the coordinates $x^0$ and $x^1$, so that $A^\mu = A^\mu(x^0,x^1)$ with $\mu = 0,1$. This restriction leads to a field strength tensor with a single independent component, $F_{01}$, whose Hodge-dual defines, up to a minus sign, the non-abelian electric field $\widetilde{F}$:
\be
{\widetilde F} \equiv - \frac{1}{2}\vareps_{\mu \nu}\,F^{\mu \nu} =  F_{01} = \partial_0 A_1 - \partial_1 A_0 + i\,e\,\sbr{A_0}{A_1}.
\lab{sec2:psft}
\ee
where we are using $\vareps_{01}=1$.

Hence, the Yang-Mills equations in $1+1$ dimensions take the form
\be
D_\mu {\widetilde F} =e\,{\widetilde J}_\mu,\qquad \mu=0,1 
\lab{sec2:em}
\ee
where $\widetilde{J}_\m = - \vareps_{\m \n} J^\n$ denotes the Hodge dual of the matter current density. Since the $F_{\m \n}$ and $J_\m$ are Lie algebra-valued quantities that under gauge transformations $g(x) \in G$, which change the gauge field by
\be
A_\mu(x) \longrightarrow g(x)\,A_\m(x)\,g^{-1}(x) + \frac{i}{e}\partial_\m g(x)\,g^{-1}(x),
\lab{sec2:gt}
\ee
transform as 
\br
F_{\mu\nu}(x) &\longrightarrow& g(x)\,F_{\mu\nu}(x)\,g^{-1}(x); \nonumber\\
J_\m(x)&\longrightarrow& g(x)\,J_{\mu}(x)\,g^{-1}(x),
\lab{fttr}
\er
so the Yang-Mills equations are gauge covariant, as they transform by conjugation with the single element $g(x)$.

In addition, the Yang-Mills equations do not guarantee a continuity equation for the matter current density $J_\m$. Instead, the continuity equation is obtained for a set of current densities $j_\m$ involving the gauge fields, i.e.,
\be
\partial_\n j^\n = 0, \qquad \qquad \qquad j^\n\equiv \( J^\n - i\, \sbr{A_\m}{F^{\m \n}}\).
\ee
Such currents lead to dynamically conserved quantities in time that are not gauge-invariant. Henceforth, they are not a suitable choice for physical observables of the theory. 

Our approach to construct physical observables in Yang-Mills in 1+1-dimensions consists in search for a conservation law in an integral formulation of the Yang-Mills equations, that is invariant under the choice of gauge. Such an approach can be obtained upon different choices of charge operators, which are equivalent under the Yang-Mills equations.

A central step in formulating the integral Yang-Mills equations is to define the \emph{flux} of the field strength $F_{\mu\nu}$, that is, a quantity that correlates to the amount of charged particles in a region of spacetime. In addition, a full gauge invariance of the flux is not necessary. It suffices that the flux transforms globally, by conjugation with a single group element evaluated at a fixed reference point. In such cases, the eigenvalues of the flux operator remain invariant under gauge transformations and can thus serve as meaningful observables. This motivates the introduction of a ``dressed'' field strength, defined by conjugating the original field strength with the holonomy associated with parallel transport from the reference point.

The field strength, constructed from the gauge connection $A_\mu$, inherits its local transformation properties. The holonomy $W$, defined as the path-ordered exponential of $A_\mu$ along a curve, is a solution of the equation
\be
\frac{dW}{d\sigma} + i\,e\,A_\mu(x(\sigma))\frac{dx^\mu}{d\sigma}\,W =0, \qquad \qquad W(\sigma_i) = W_R
\lab{sec2:holeq}
\ee
where $\sigma \in [\sigma_i,\sigma_f]$ parameterizes the path in space-time, such that $x(\sigma_i) \equiv x_R$ is the initial point or ``reference point'', implements parallel transport and encapsulates the gauge principle in geometric terms. It encodes how internal degrees of freedom are transported along space-time paths and is locally transformed under gauge transformations:
\be
W \longrightarrow g (x)\; W\;{\widetilde g}^{-1}(x_R), \qquad \qquad \qquad {\widetilde g}(x_R) \equiv W_R^{-1}\,g(x_R)\,W_R
\lab{sec2:wgt}
\ee
where $W_R$ is an integration constant, an element of the gauge group $G$, which does not transform under the gauge transformations. 

By conjugating the field strength as $F^W_{\mu\nu}\equiv W^{-1} F_{\mu\nu} W$, we effectively translate it to the reference point $x_R$, where gauge transformations act only globally, that is,
\be
W^{-1}F_{\mu\nu} W\ \rightarrow \wti{g}(x_R)\,W^{-1}\,F_{\mu\nu}\,W\,\wti{g}^{-1}(x_R)
\ee
The local character of the transformation is absorbed into the holonomy, and the resulting conjugated field transforms as a Lie-algebra element under a single global group action. This construction provides a gauge-covariant mechanism for coupling the field strength to the gauge structure via the holonomy, and allows for the definition of non-local but gauge-invariant observables, such as the spectrum of the flux operator.

In 2-dimensional space-time, the usual geometric interpretation of flux as the integral of a field over a surface must be adapted to the reduced dimensionality: a ``closed surface'' consists of a discrete set of two points in space-time. At the same time, the ``volume'' becomes a one-dimensional path connecting them. The flux of the non-abelian electric field ``across the surface'' $x$ is therefore defined by its value at that point, i.e., $\wti{F}(x)$. Hence, if we define the flux of the conjugated non-abelian electric field as follows 
\be
\Phi(x)\equiv ie\beta W^{-1} \widetilde{F} W(x),
\lab{sec2:phi}
\ee
with $\beta$ an arbitrary constant, it transforms under a global gauge transformation at the reference point $x_R$. Note, at the cost of removing the local gauge action, constructing a non-local flux \rf{sec2:phi} we get, instead, the freedom to choose any path with endpoint $x$ where the Wilson line is integrated.

We can consider the total flux of the field across a ``closed surface'' $\{x_R \cup x\}$ is given by the difference in its values at these two points:
\be
\Delta \Phi = \Phi(x)-\Phi(x_R) = ie\beta\left(W^{-1} \widetilde{F} W(x) - W^{-1}_R\widetilde{F}(x_R)W_R\right).
\lab{sec2:totflux}
\ee
where $W_R= W(x_R)$ denotes the Wilson line evaluated at the reference point.

Alternatively, this same total flux can be obtained by transporting the conjugated field from the reference point $x_R$ to the point $x$ along a one-dimensional path and accumulating infinitesimal variations along the way. That is, we consider how the quantity $\Phi(x)$ changes under a small displacement $x \to x + \delta x$, and integrate this variation along the path connecting $x_R$ to final point: 
\br
\delta \Phi &=& ie\beta \left(\delta W^{-1}\,\widetilde{F}\,W +W^{-1}\delta \widetilde{F} \,W + W^{-1}\,\widetilde{F}\,\delta W\right)\nonumber\\
&=&ie\beta \left(W^{-1}\delta \widetilde{F}W + [W^{-1}\widetilde{F}W,W^{-1}\delta W]\right)=ie\beta W^{-1}\left[\partial_\mu \widetilde{F} + ie\left[A_\mu,\widetilde{F}\right]\right]W\frac{dx^\mu}{d\sigma}\delta \sigma\nonumber\\
&=&ie\beta W^{-1}D_\mu \widetilde{F}W\frac{dx^\mu}{d\sigma}\delta \sigma.
\lab{dphi}
\er
where we have used \rf{sec2:holeq} to write $\delta W = -ieA_\mu W \delta x^\mu$. Finally, with $\delta \Phi = \frac{d\Phi}{d\sigma}\delta \sigma$, one obtains a differential equation that gives the change of flux from point to point along the path
\be
\frac{d\Phi}{d\sigma} = ie\beta W^{-1}D_\mu \widetilde{F}W\frac{dx^\mu}{d\sigma}
\ee
which, after integration from $x_R = x(\sigma_i)$ to $x = x(\sigma)$, results in the total flux across the borders of the path
\be
\Delta \Phi = \int_{\sigma_i}^\sigma d\sigma'\,W^{-1}D_\mu \widetilde{F}W\frac{dx^\mu}{d\sigma'}.
\lab{sec2:fundt0}
\ee
The fact that the total flux can be computed either directly from the difference in values of the conjugated field at the endpoints or from the integral of the conjugated covariant derivative of the electric field along the path, as in \rf{sec2:fundt0}, is the identity:
\begin{equation}
W^{-1} \widetilde{F} W(x) - W^{-1}_R\widetilde{F}(x_R)W_R = \int_{\sigma_i}^{\sigma} d\sigma' \, W^{-1} D_\mu \widetilde{F} \, W \, \frac{dx^\mu}{d\sigma'}.
\lab{sec2:fundt}
\end{equation}
If the Yang-Mills equations \rf{sec2:em} yield, we can substitute them in the right-hand side of the identity \rf{sec2:fundt} and obtain an integral dynamical equation
\begin{equation}
W^{-1} \widetilde{F} W(x) - W^{-1}_R\widetilde{F}(x_R)W_R = e\int_{\sigma_i}^{\sigma} d\sigma' \, W^{-1} \wti{J}_\m \, W \, \frac{dx^\mu}{d\sigma'}.
\lab{inteqfail}
\end{equation}

Although equation \rf{sec2:fundt} provides an integral version of the dynamical equations and can be used to construct conserved quantities for the system, it is known that the Yang-Mills theories can possess an infinite hierarchy of dynamical equations and conserved and gauge-invariant quantities \cite{lafgl1,lafgl2}, and such quantities lead to novel symmetries in the Hamiltonian formalism \cite{ymintg}. Thus, using the integral equation \rf{inteqfail} to construct physical observables fails to reveal hidden structures in (1+1)-dimensional Yang-Mills theory.  

Our approach to deriving the conserved and gauge-invariant charges relies on integrability techniques, constructing a set of path-ordered operators, called holonomies, that depend on a complex(real) spectral parameter. The conservation law is expressed by a path-invariance condition on such operators, which may have a zero curvature representation. Such structures are naturally formulated on loop spaces, and the dynamics in the loop space are expressed in terms of group elements, as we do so next.

\subsection{The integral formulation of the dynamical equations in terms of holonomies}
In two-dimensional Minkowski spacetime, the loop space $\mathcal{L}^{(0)}(\mathcal{M})$ is defined as the space of maps from $0$-dimensional closed submanifolds into spacetime:
\be
\mathcal{L}^{(0)}(\mathcal{M}) \equiv \{f : S^0 \rightarrow \mathcal{M} \ | \ f(-1) = x_R, \ x_R \in \mathcal{M}\}
\ee 
where $S^0$ is the zero-sphere, given by $z^2 = 1$, with $z=\pm 1$. Since $x_R$ is a fixed base point in $\mathcal{M}$, the image of each map $f \in  \mathcal{L}^{(0)}(\mathcal{M})$ at $z=1$ corresponds to a different point $x \in \mathcal{M}$, and thus we can conclude that point of the loop space $\mathcal{L}^{(0)}(\mathcal{M})$ can be mapped in a one-to-one correspondence to the points of the spacetime. In this context, each ``loop'' corresponds to a single point in spacetime, and thus the loop space is naturally identified with the base manifold. As a result, the loop space formalism becomes a point-based description, while still retaining the non-local structure of parallel transport through the holonomy operator defined along paths. This loop space holonomy, that is, the parallel transport operator in the loop space, is defined as the ``flux operator''
\be
V (\b;x) \equiv e^{ie \beta W^{-1} \widetilde{F} W(x)}
\lab{vdef}
\ee
and we derive the corresponding Stokes theorem for it by noticing that under an infinitesimal variation of the space-time point, the variation of $V$ is given by
\be
\delta V\,V^{-1} = \delta \Phi+\frac{1}{2}[\Phi,\delta \Phi]+\frac{1}{3!}[\Phi,[\Phi,\delta \Phi]] +\dots
\lab{sec2:deltav}
\ee
with $\Phi$ defined in \rf{sec2:phi}. Introducing the operator for the adjoint action of $\widetilde{F}$ as
\be
L_\mu \equiv \sum_{n=0}^{\infty}\frac{(ie\beta)^{n+1}}{(n+1)!}\textrm{ad}_{\widetilde{F}}^n D_\mu\widetilde{F} =ie\beta\, D_\mu \widetilde{F}+\frac{(ie\beta)^{2}}{2}[\widetilde{F},D_\mu\widetilde{F}]+\frac{(ie\beta)^{3}}{3!}[\widetilde{F},[\widetilde{F},D_\mu\widetilde{F}]]+\dots
\lab{sec2:lfaux}
\ee
and considering a path $x(\sigma)$ with $\sigma \in [\sigma_i, \sigma_f]$, such that for any quantity $f(x)$, its infinitesimal change along the path is given by $\delta f = \delta \sigma \frac{df}{d\sigma}$, the expression \rf{sec2:deltav}, using \rf{dphi}, defines the differential equation
\be
\frac{dV}{d\sigma}-\left(W^{-1}L_\mu W\right) \frac{dx^\mu}{d\sigma}\,V = 0
\lab{sec2:v_eq}
\ee
which can be solved iteratively, resulting in the path-ordered series
\be
V_\G = P\,e^{\int_{\Gamma}d\sigma \,W^{-1}L_\mu W\,\frac{dx^\mu}{d\sigma}}\,V_R
\lab{sec2:vg}
\ee
the flux operator at the reference point and $\Gamma$, the curve joining $x_R\equiv x(\sigma_i)$ and $x(\sigma_f)$, with $V_R$ an integration constant.
For the sake of notation, the series in \rf{sec2:lfaux} can be represented in terms of an integral over an auxiliary real parameter $\l \in [0,1]$, such as,
\be
L_\mu = i\,e\,\b \int_{0}^{1}d\l\,{\rm e}^{i\,e\,\b\,\l\,\wti{F}}\,D_\m \wti{F}\,{\rm e}^{-i\,e\,\b\,\l\,\wti{F}}.
\lab{sec2:lf}
\ee

The result \rf{sec2:vg} shows that $V$ can be evaluated either directly, via the exponential of the electric field, i.e., as in the definition \rf{vdef}, or via path-ordered integration, leading to the mathematical identity:
\be
e^{ie\beta W^{-1}\widetilde{F}W} = P\,e^{\int_{\Gamma}d\sigma \,W^{-1}L_\mu W\,\frac{dx^\mu}{d\sigma}}\,V_R.
\lab{sec2:hint}
\ee
In addition, it fixes the value of $V_R$ as 
\be
V_R = e^{ie\beta W_R^{-1}\widetilde{F}(x_R)W_R}.
\lab{sec2:vr}
\ee

We formulate the integral Yang-Mills equations in terms of elements of the loop space by using the relation \rf{sec2:hint} and by imposing the local dynamical equations (e.o.m.) \rf{sec2:em} in the connection $L_\mu$. Thus, we define the new connection as
\be
K_\m(\b,x) \equiv i\,e^2\,\b \int_{0}^{1}d\l\,{\rm e}^{i\,e\,\b\,\l\,\wti{F}(x)}\,\wti{J}_\m(x)\,{\rm e}^{-i\,e\,\b\,\l\,\wti{F}(x)}  
\lab{sec2:lj}
\ee
which is equivalent to $L_\m$ when the equations of motion \rf{sec2:em} yields.

One can construct a holonomy associated to the non-local connection $W^{-1}K_\m W$ as a solution of a ordinary differential equation, along a path $\G$, described by the chart $x(\s)$, with $\s \in [\s_i, \s_f]$, and $x(\s_i) = x_R$, given by
\be
\frac{dU_{x_R}(\b,\s)}{d\s} -W_{\G}^{-1}(\s)\,K_\m(\b,x(\s))\,W_\G(\s) \, U_{x_R}(\b,\s)\,\frac{dx^\m}{d\s} = 0, \qquad  U_{x_R}(\b,\s_i) = U_R(\b)
\lab{holeqq}
\ee
where $U_R(\b)$ is an integration constant that may depends on $\b$.

Denoting by ${\cal Q}_{x_R}$, the path-ordered exponential, solution of the holonomy equation \rf{holeqq} which is close to the identity when $\s = \s_i$, as follows
\be
{\cal Q}_{x_R}(\G) = P\,e^{\int_{\Gamma}d\sigma \,W_\G^{-1}K_\mu W_\G\,\frac{dx^\mu}{d\sigma}}.
\ee
the general solution for \rf{holeqq}, will be
\be
U_{x_R}(\G) = {\cal Q}_{x_R}(\G) U_R.
\lab{opdef}
\ee
where we omitted the dependence on $\b$, i.e., $Q_{x_R}(\b,\s)= Q_{x_R}(\s)$,
and the subscript $x_R$ in ${\cal Q}_{x_R}$ denotes the explicit dependence of the operator upon the reference point $x_R$ where the Wilson line starts. In fact, the reference point can be taken independently of the path $\G$ where ${\cal Q}_{x_R}(\G)$ is integrated. The changing of the reference point from $x_R$ to $x^\prime_R$, keeping $\G$ fixed, is dictated by the decomposition law of holonomies of the Wilson lines along the path that joins the two reference points, and it leads to the transformation
\be
{\cal Q}_{x_R^\prime}(\G) = W(x_R \rightarrow x_R^\prime){\cal Q}_{x_R}(\G)W^{-1}(x_R \rightarrow x_R^\prime)
\lab{rpc}
\ee
Such a property is directly related to the fact that ${\cal Q}$ is a holonomy in the ${\cal L}^{(0)}$ loop space.

When the Yang-Mills equations, i.e., the equations of motion (e.o.m.) expressed in \rf{sec2:em}, are imposed, the connections $L_\m$ and $K_\m$ are found to be equal. Hence, from the identity \rf{sec2:hint}, we can obtain an integral formulation of local Yang-Mills equations:
\be
V(\b,\G)V_R^{-1}(\b) = e^{ie\beta W_\G^{-1}\widetilde{F}W_\G}e^{-ie\beta W_R^{-1}\widetilde{F}(x_R)W_R} \overset{e.o.m.}{=}P\,e^{\int_{\Gamma}d\sigma \,W_\G^{-1}K_\mu W_\G\,\frac{dx^\mu}{d\sigma}}= {\cal Q}_{x_R}(\b,\G),
\lab{sec2:hintaux0}
\ee
or equivalently
\be
    V(\b,\G) = e^{ie\beta W_\G^{-1}\widetilde{F}W_\G} \overset{e.o.m.}{=}P\,e^{\int_{\Gamma}d\sigma \,W_\G^{-1}K_\mu W_\G\,\frac{dx^\mu}{d\sigma}}U_R(\b)= U_{x_R}(\b,\G)
\lab{sec2:hintaux}
\ee 
with $U_{x_R}(\G)$ given by \rf{opdef}. The integration constant $U_R$ is identified with $V_R$ through the integral equations \rf{sec2:hintaux}, i.e.,
\be
V(\b,x_R) = V_R =e^{ie\beta W_R^{-1}\widetilde{F}(x_R)W_R} \overset{e.o.m.}{=} U_R(\b).
\lab{urvr}
\ee

The rearranged integral equations \rf{sec2:hintaux} retain an important property that the eigenvalues of both sides of the equation are independent of the path $\G$ as long as their end points are kept fixed. Such a property can be verified by considering a path $\G^\prime$ with the same endpoints of $\G$, such that the combined path $\G \circ \G^{\prime^{-1}}$  is a closed path. The Stokes theorem for a $1$-form connection \cite{lafgl1, afg1, afg2} states the following
\be
W(\G^{\prime^{-1}})\,W(\G) = P_2e^{-ie\oint_{\S} d\s d\t \,W^{-1}\,F_{\m \n}\,W\, \frac{dx^\m}{d\s} \frac{dx^\n}{d\t} } = H(\S)
\ee
where $\S$ is any surface with boundary $\partial \S = \G \circ \G^{\prime^{-1}}$, scanned by loops, based at the reference point $x_R$, denoted by $\t$ and parameterized by $\s$. Hence, it follows that
\be
V(\G^\prime) = H(\S)\,V(\G)\,H^{-1}(\S).
\lab{vH}
\ee
So, under the change of paths from $\G$ to $\G^\prime$, the operator in the l.h.s of the equation \rf{sec2:hintaux} transforms by the conjugation of $H(\S)$, and as a result, their eigenvalues are path independent. In addition to that, the integral equation \rf{sec2:hintaux} assures the same property for the r.h.s., and, since the eigenvalues of a matrix can be written functionally as the trace of its powers, one can write a path-independent integral equation
\be
{\rm Tr}\,V^N(\G) \overset{e.o.m.}{=} {\rm Tr}\,U_{x_R}^N(\G) ={\rm Tr}\({\cal Q}_{x_R}(\G)U_R\)^N
\lab{sec2:hinteq}
\ee

Furthermore, both sides of the equation \rf{sec2:hintaux} can be expanded in a series of the $\beta$ parameter if we consider that $U_R(\b) = \sum_{n=0}^{+\infty}U^{(n)}_R\,\b^n$ . Since the parameter is arbitrary, taking values over the real (complex) line, for each order of $\beta$ there is a dynamical equation, relating the coefficients of both sides of the equation \rf{sec2:hintaux}, that is, in zero order in $\beta$:
\be
1 \overset{e.o.m.}{=} U_{R}^{(0)}.
\ee
In first order in $\beta$ 
\br
{\widetilde F}^W(\s_f) \overset{e.o.m.}{=}  e\int_{\s_i}^{\s_f} d\s {\widetilde J}^W_\m\frac{dx^\m}{d\s} \,U_R^{(0)}+\frac{1}{ie}U_R^{(1)}
\lab{feq}
\er
In the second order of $\beta$:
\br
&&\frac{1}{2!}\({\widetilde F}^W(\s) \)^2  \overset{e.o.m.}{=} \frac{1}{(ie)^2}U_R^{(2)} + e\(\int_{\s_i}^{\s_f} d\s {\widetilde J}^W_\m\frac{dx^\m}{d\s}\)U^{(1)}_R \\
&&+\(\frac{e}{2!}\int_{\s_i}^{\s_f} d\s \sbr{{\widetilde F}}{{\widetilde J}_\m}^W\,\frac{dx^\m}{d\s} + e^2\int_{\s_i}^{\s_f} d\s \int_{\s_i}^{\s} d\s^\prime\,{\widetilde J}_\m^W(\s){\widetilde J}_\n^W(\s^\prime) \frac{dx^\m}{d\s}\frac{dx^\n}{d\s^\prime}\)U_R^{(0)}\nonumber
\lab{seq}
\er
In the third order of $\beta$:
\br
&&\frac{1}{3!}\({\widetilde F}^W(\s)\)^3  \overset{e.o.m.}{=} \frac{1}{(ie)^3}U_R^{(3)}
+ e\(\int_{\s_i}^{\s_f} d\s {\widetilde J}^W_\m\frac{dx^\m}{d\s}\)U_R^{(2)} \\
&&+\(\frac{e}{2!}\int_{\s_i}^{\s_f} d\s \sbr{{\widetilde F}}{{\widetilde J}_\m}^W\,\frac{dx^\m}{d\s} + e^2\int_{\s_i}^{\s_f} d\s \int_{\s_i}^{\s} d\s^\prime\,{\widetilde J}_\m^W(\s){\widetilde J}_\n^W(\s^\prime) \frac{dx^\m}{d\s}\frac{dx^\n}{d\s^\prime}\) U_R^{(1)}\nonumber\\
&&+\bigg(\frac{e}{3!}\int_{\s_i}^{\s_f} d\s \sbr{{\widetilde F}}{\sbr{{\widetilde F}}{{\widetilde J}_\m}}^W\,\frac{dx^\m}{d\s} +e^3\int_{\s_i}^{\s_f} d\s \int_{\s_i}^{\s} d\s^\prime \int_{\s_i}^{\s^\prime} d \s^{\prime \prime} {\widetilde J}^W_\m(\s){\widetilde J}^W_\n(\s^\prime){\widetilde J}^W_\l(\s^{\prime \prime}) \nonumber\\
&&+\frac{e^2}{2!}\int_{\s_i}^{\s_f} d\s \int_{\s_i}^{\s} d\s^\prime \(\sbr{\widetilde F}{{\widetilde J}_\m}^W(\s)\, {\widetilde J}_\n^W(\s^\prime)+{\widetilde J}_\n^W(\s)\,\sbr{\widetilde F}{{\widetilde J}_\n}^W(\s^\prime)\)\frac{dx^\m}{d\s}\frac{dx^\n}{d\s^\prime}\bigg)U_R^{(0)}\nonumber\\
\nonumber
\lab{teq}
\er
and so on.
The superscript $W$, as $X^W$, denote the conjugation by Wilson line $X^W=W^{-1} X W$. Note that, in fact, we have
\be
U^{(n)}_R \overset{e.o.m.}{=}\frac{(i\,e)^n}{n!}\(W_R^{-1}\wti{F}(x_R)W_R\)^n.
\ee
that is $U_R(\b) \overset{e.o.m.}{=}V_R(\b)$.
Hence, the equation \rf{inteqfail} corresponds to the first-order equation in the $\b$ expansion, i.e., \rf{feq}. In fact, it solves all the equations in the hierarchy when the Yang-Mills equations are satisfied. Still, the infinite quantities obtained in the right-hand side of the $\b$ expansion in \rf{sec2:hintaux} have different effects in Hamiltonian formalism as we shall see later, which cannot be observed if we do not construct a path-ordered operator as $U_{x_R}(\b, \G)$.

Moreover, the formulation of the integral equation can be, in fact, arbitrary. There are multiple ways to construct an operator $U_{x_R}(\b,\G)$, all equivalent when the Yang-Mills equations are satisfied, with different properties otherwise. The definition \rf{vdef} is motivated by an operator that has global gauge transformations. Such a condition is achieved by parallel transport to a fixed reference point through Wilson line conjugation. Hence, such an operator has global gauge transformations. 

Instead of defining it through the Wilson-line conjugation in \rf{vdef}, we take $V \equiv \exp(ie\b \wti{F})$. This definition can be associated with a holonomy $U_{x_R}(\b,\Gamma)$ through the corresponding integral equations, which define a locally flat connection on space-time when the Yang–Mills equations are satisfied. However, the resulting holonomy $U_{x_R}(\b,\Gamma)$ does not transform regularly under gauge transformations unless the Yang–Mills equations hold true.

\subsection{The non-abelian dynamically conserved charges}

The expression \rf{sec2:hinteq} explicitly exhibits the fundamental property of path-independence: with $x$ kept fixed, the result is independent of the chosen path $\Gamma$, with endpoint $x$. In the integral formulation, this geometric feature becomes a conservation law. It ensures that isospectral evolution can be described by composing holonomies over different paths.

The path-independence guaranteed from the integral equations in the form of \rf{sec2:hinteq} yields, for two different paths $\Gamma $ and $\Gamma^\prime$ with the same initial and final points, the following relation
\be
 {\rm Tr}\({\cal Q}_{x_R}(\G^\prime)V_R\)^N = {\rm Tr}\({\cal Q}_{x_R}(\G)V_R\)^N,
\lab{sec2:pic}
\ee
where we considered that \rf{urvr}, since the path-invariance is guaranteed only when the Yang-Mills equations yield.

\begin{figure}[htbp]
    \centering
    \begin{tikzpicture}[
        >=latex, 
        mid arrow/.style={postaction={decorate, decoration={
            markings,
            mark=at position 0.55 with {\arrow{>}}
        }}},
        xscale=1.5, 
        yscale=1.5
    ]

    \begin{scope}
        \def\L{1.5} 
        \def\T{1.2} 

        \draw[->] (-\L-0.5, 0) -- (\L+0.5, 0) node[right] {$x$};
        \draw[->] (0, 0) -- (0, \T+0.8) node[above] {$t$};

        \draw (-\L, 0.1) -- (-\L, -0.1) node[below] {$-L$};
        \draw (\L, 0.1) -- (\L, -0.1) node[below] {$L$};

        \draw[thick, mid arrow] (-\L, \T) -- (\L, \T) node[pos = 0.4, above=2pt] {$\Gamma_t$};

        \node[circle, fill, inner sep=1.5pt, label=$x_R^{(t)}$] at (-\L, \T) {};
        
    \end{scope}

    \begin{scope}[xshift=5cm] 
        \def\L{1.5}
        \def\T{1.2}

        \draw[->] (-\L-0.5, 0) -- (\L+0.5, 0) node[right] {$x$};
        \draw[->] (0, 0) -- (0, \T+0.8) node[above] {$t$};

        \draw (-\L, 0.1) -- (-\L, -0.1) node[below] {$-L$};
        \draw (\L, 0.1) -- (\L, -0.1) node[below] {$L$};

        \draw[thick, mid arrow] (-\L, \T) -- (-\L, 0) node[midway, left] {$\Gamma_{-L}^{-1}$};

        \node[circle, fill, inner sep=1.5pt, label=$x_R^{(t)}$] at (-\L, \T) {};

        \draw[thick, mid arrow] (-\L, 0) -- (\L, 0) node[midway, below=5pt] {$\Gamma_0$};

        \draw[thick, mid arrow] (\L, 0) -- (\L, \T) node[midway, right] {$\Gamma_L$};
    \end{scope}

    \end{tikzpicture}
    
    \caption{Two paths in $(1+1)$-spacetime $\Gamma = \Gamma_t$ and $\Gamma' = \Gamma_{-L}^{-1} \circ \Gamma_0 \circ \Gamma_L$, with the same initial point $x^\mu = x_R^{(t)} = (t, -L)$ and final point $x^\mu = (t, +L)$.}
    \label{fig:twopaths}
\end{figure}
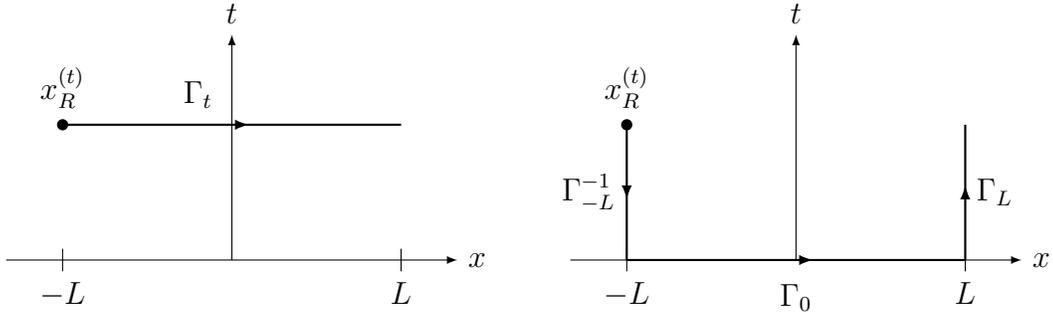

Now consider the specific paths $\Gamma = \Gamma_t$ and $\Gamma' = \Gamma_{-L}^{-1} \circ \Gamma_0 \circ \Gamma_L$, as illustrated in Figure~\ref{fig:twopaths}, going from $(t, -L)$ to $(t, +L)$. The composition $\G \circ \G^{\prime \, ^{-1}}$ describes a closed square in the two-dimensional Minkowski spacetime. As ${\cal Q}_{x_R^{(t)}}(\G^\prime)$ satisfies the holonomy equation \rf{holeqq}, fixing the reference point $x_R^{(t)}$, the operator follows the decomposition law of holonomies along the composed paths $\G^\prime =\Gamma_{-L}^{-1} \circ \Gamma_0 \circ \Gamma_L$ so that we can write the following:
\br
{\cal Q}_{x_R^{(t)}}(\Gamma^\prime) &=& P\,e^{\int_{\Gamma_{L}}d\s\,\frac{dx^\m}{d\s} \,W_{\G^\prime}^{-1}K_\m(t,+L) W_{\G^\prime}}\,P\,e^{\int_{\Gamma_0}d\s\,\frac{dx^\m}{d\s} \,W_{\G^\prime}^{-1}K_\m(0, x) W_{\G^\prime}}\,P\,e^{\int_{\Gamma_{-L}^{-1}}d\s\,\frac{dx^\m}{d\s} \,W_{\G^\prime}^{-1}K_\m(t,-L) W_{\G^\prime}}\nonumber\\
&=& {\cal Q}_{x_R^{(t)}}(\G_L)\,{\cal Q}_{x_R^{(t)}}(\Gamma_0)\,{\cal Q}_{x_R^{(t)}}(\G_L^{-1}).
\lab{sec2:qp}
\er
Note that the holonomies in the r.h.s. of \rf{sec2:qp} share the same reference point of ${\cal Q}_{x_R^{(t)}}$, which is not the start point of the paths where they are integrated. In order to get the holonomies at the start point of each path, we use the property \rf{rpc}, such that one can write \rf{sec2:qp} as follows
\br
&&{\cal Q}_{x_R^{(t)}}(\Gamma^\prime) =\lab{qp1}\\
&&=W^{-1}(\G_{-L}^{-1} \circ \G_0){\cal Q}_{(0,+L)}(\G_L)W(\G_{-L}^{-1} \circ \G_0)\,W^{-1}(\G_{-L}^{-1}){\cal Q}_{x_R^{(0)}}(\Gamma_0)W(\G_{-L}^{-1})\,{\cal Q}_{x_R^{(t)}}(\G_L^{-1})
\nonumber
\er

Now, considering the following boundary conditions for the matter fields
\be
J_1(t, \pm L) \rightarrow \frac{1}{L^\d}; \qquad   {\rm as}  \qquad L \rightarrow +\infty, \qquad \qquad {\rm and}\qquad  \d > 0
\lab{sec2:bc1}
\ee
on the expression \rf{sec2:qp},
since $K_\mu$ is given in terms of the matter currents (see \rf{sec2:lj}) and $\widetilde{J}_0 = J_1$, we get from \rf{sec2:lj}, that
\be
{\cal Q}_{(0,+\infty)}(\G_\infty) =\mathbb{I}, \qquad \qquad {\cal Q}_{x_R^{(t)}}(\G_\infty^{-1})= \mathbb{I} 
\ee
and then, the expression \rf{qp1} becomes
\be
{\cal Q}_{x_R^{(t)}}(\Gamma^\prime) =W^{-1}(\G_{-\infty}^{-1}){\cal Q}_{x_R^{(0)}}(\Gamma_0)W(\G_{-\infty}^{-1})
\ee
Substituting it into the path-independent relation \rf{sec2:pic}, one gets that
\be
{\rm Tr}\({\cal Q}_{x_R^{(t)}}(\Gamma_t)V_R^{(t)}\)^N ={\rm Tr}\(W^{-1}(\G_{-\infty}^{-1}){\cal Q}^N_{x_R^{(0)}}(\Gamma_0)W(\G_{-\infty}^{-1})V_R^{(t)}\)^N
\lab{sec2:relgpg1}
\ee
If we impose the following boundary conditions for the time components of the gauge fields
\be
A_0(t, -L)  \rightarrow \frac{1}{L^{\d^\prime}}; \qquad   {\rm as}   \qquad  L \rightarrow +\infty, \qquad  {\rm and}  \qquad  \d^\prime > 0
\lab{sec2:bc2}
\ee
we get that
\be
W(\G_{-\infty}^{-1}) = W_R .
\ee
Note yet, from \rf{sec2:em}, \rf{sec2:bc1} and \rf{sec2:bc2}
we have that
\be
\partial_0 \wti{F}(t,-L) \rightarrow 0 \qquad {\rm as} \qquad L \rightarrow \infty
\ee
Consequently $V_R^{(t)}$ becomes a constant element of the gauge group $G$, since $\wti{F}(t,-\infty)$ becomes a constant value for any time $t$ at $x = -\infty$. Therefore, we get that
\be
{\rm Tr}\({\cal Q}_{x_R^{(t)}}(\Gamma_t)V_R\)^N ={\rm Tr}\({\cal Q}_{x_R^{(0)}}(\Gamma_0)W_R V_R W_R^{-1}\)^N  = 
{\rm Tr}\({\cal Q}_{x_R^{(0)}}(\Gamma_0)e^{ie\b\wti{F}(-\infty)}\)^N
\lab{sec2:relgpg2}
\ee
note we omitted the time dependence in $\wti{F}(t,-\infty) = \wti{F}(-\infty)$, as long as we fix the time $t$.
As we have shown in \rf{sec2:vr}, the value of $V_R$ depends on the integration constants of $W_R$ at the reference point $x_R^{(t)}$, the initial point of $\G_t$. The initial point of $\G_0$ is $(t=0, -\infty)$. So, if we instead of considering $W_R$ as the integration constant at $x_R^0$, consider $W_{R_0}$, we then have $V_{R_0}= e^{ie\b W_{R_0}^{-1}\wti{F}(-\infty)W_{R_0}}$, and \rf{sec2:relgpg2} can becomes:
\be
{\rm Tr}\({\cal Q}_{x_R^{(t)}}(\Gamma_t)V_R\)^N  = {\rm Tr}\({\cal Q}_{x_R^{(0)}}(\Gamma_0)V_{R_0}\)^N 
\ee
From \rf{opdef}, one concludes that the quantities
\be
q_N \equiv \frac{1}{N}{\rm Tr}\(Q^N_{x_R^{(t)}}(t,\b)\)
\lab{sec2:cc}
\ee
are conserved in time. From the gauge transformations \rf{fttr} and \rf{sec2:wgt}, one can see that the charge operator $Q_{x_R^{(0)}}$ has a global transformation at fixed $x_R^{(0)}$:
\be
Q_{x_R^{(0)}} \longrightarrow \wti{g}(x_R^{(0)})\,Q_{x_R^{(0)}}\,\wti{g}^{-1}(x_R^{(0)}), \qquad \qquad g(x_R^{(0)}) \in G
\ee
so, the conserved charges \rf{sec2:cc} are also gauge-invariant. Thus, the quantities \rf{sec2:cc} are suitable candidates as physical observables for non-abelian Yang-Mills theories in two dimensions, as well as the holonomy \rf{opdef} as the charge operator when evaluated along the real axis $\G=\mathbb{R}$. Whether these charges can represent new quantum numbers in Yang-Mills theory requires further analysis, but it has been shown that in two-dimensional lattice QCD, matter states (barions and mesons) can carry these charges \cite{twolattice}.

Moreover, the charge operator \rf{opdef} shares the same expression of monodromies in two-dimensional integrable models \cite{beisert2014introduction, retore2022introduction}. The parameter $\b$ can be understood as a spectral parameter of the charge operator. In addition, one can perform a series expansion of \rf{opdef} on the arbitrary parameter $\b$, which, in principle, leads to an infinite number of conserved quantities, eigenvalues of operators that are coefficients of the following expansion
\be
Q_{x_R^{(t)}}(\b, t) = \sum_{n=0}^{+\infty} Q_{(n)}(t)\b^n. 
\lab{chexp}
\ee
This repeats the structure of an infinite tower of conserved charges in $(3+1)$-dimensional Yang-Mills theories \cite{lafgl1, lafgl2, ymintg}. 

\section{The symplectic structure}
\label{sec:3}
\setcounter{equation}{0}
Having identified a set of gauge-invariant conserved charges from the integral formulation of Yang-Mills theory, we now investigate how these charges generate symmetry transformations on the physical fields. To this end, we turn to the canonical formalism, where conserved quantities are naturally interpreted as generators of flows on phase space via the symplectic structure.

In particular, the Poisson brackets between the conserved charges and the fundamental fields encode how the charges act dynamically. This requires a precise formulation of the phase space and its symplectic structure, which is best achieved in the first-order (or symplectic) formalism.

We adopt the approach developed in \cite{faddeevslavnov}, where the Yang-Mills theory is described in first-order form in ($3+1$)-dimensional Minkowski spacetime. This formulation is particularly well-suited to canonical analysis, as the field strength $F_{\mu\nu}$ is treated as an independent auxiliary field, simplifying the identification of canonical variables.

In such an approach, the action of the model reads
\br
{\cal S}_{FO}&=& \int d^2x\;\Bigg\{ -\frac{1}{2}\, {\rm Tr}\left[\(\partial_{\mu}A_{\nu}-\partial_{\nu}A_{\mu}+i\,e\,\sbr{A_{\mu}}{A_{\nu}}-\frac{1}{2}\,F_{\mu\nu}\)F^{\mu\nu}\right]\nonumber\\
&+& {\bar \psi}^{(f)}\(i\,\gamma^{\mu}\,D_{\mu}-M_{(f)}\)\psi^{(f)} +\vp_\m^\dagger D^{\mu}\vp + \(D^\m \vp\)^{\dagger}\vp_\m -V\(\mid \vp\mid\) - \vp^\dagger_\m \vp^\m\Bigg\}
\lab{sec4:ymlagfirstorder}
\er
where the covariant derivatives are given by
\br
D_\mu \psi^{(f)} &=& \partial_\mu \psi^{(f)} + i\,e\,R_{\psi}(A_\mu)\,\psi^{(f)},\nonumber\\
D_\mu \vp &=& \partial_\mu \vp + i\,e\,R_{\phi}(A_\mu)\,\vp.
\er
where $\vp_\m$ and $\vp_\m^\dagger$ are auxiliary variables. 
The fields $\psi^{(f)}$ with flavor index $f$ representing the fermions, and scalar fields $\varphi$ represent the bosons. The fields $A_\mu$ take values in the Lie algebra of the gauge group $G$, and $R_\psi$, $R_\phi$ denote the representations of $G$ under which the fermions and bosons transform, respectively.

In the first-order formalism, the fields $\psi$, $\psi^\dagger$, $\varphi$, $\varphi^\dagger$, $\vp_\m$, $\vp_\m^\dagger$, $A_1$, and $F_{01}$ are treated as independent variables. The gamma matrices $\gamma_\mu$ are chosen to be anti-Hermitian, and the spinor mass matrix is taken as $M_{(f)} = m\,\delta_{AB}$. The covariant derivative acts as $D_\mu = \partial_\mu + i\,e\,[A_\mu,\,\cdot\,]$, with the gauge fields $A_\mu$ valued in the Lie algebra of the gauge group $G$, i.e., $A_{\mu} = A_{\mu}^a\,T_a$. In $1+1$ dimensions, there is no magnetic field, as there is only one spatial direction. The only non-zero component of the field strength tensor is $F_{01}$, which can be dualized into the pseudoscalar quantity $\widetilde{F}$, as introduced in equation \rf{sec2:psft}.

From the action \rf{sec4:ymlagfirstorder}, we compute the canonical momenta. For the gauge field $A_1^a$, we find
\be
\pi^a = \frac{\delta S_{FO}}{\partial(\partial_0 A_1)^a} = {\widetilde F}^a,
\lab{sec4:gaugemoment}
\ee
and for the spinor field $\psi$, the momentum is
\be
\pi^{\psi} = \frac{\delta S_{FO}}{\delta (\partial_0 \psi)} = i\,\psi^\dagger,
\lab{sec4:spinormoment}
\ee
where we suppress representation and spinor indices for clarity. For the boson field $\varphi$ and its adjoint, we obtain:

\be
\pi_\vp = \frac{\delta S_{FO}}{\delta (\partial_0 \vp)} = \vp_0^\dagger;\qquad\qquad
\pi_{\vp^{\dagger}}=\frac{\delta S_{FO}}{\delta\,\partial_0\,\vp^{\dagger}}= \vp_0.
\lab{sec4:bosonmoment}
\ee

Up to a total derivative, the Lagrangian density in the action \rf{sec4:ymlagfirstorder} can then be expressed in the canonical form
 \be
{\cal L}= \widetilde{F}^a\,\partial_0A_1^a+\pi_A^\psi\,\partial_0\psi_{A}+\pi_\vp\,\partial_0\vp+\pi_{\vp^\dagger}\,\partial_0\vp^{\dagger} -{\cal H}+A_0^a\,{\cal C}_a,
\lab{sec4:legt}
\ee
where $A_0^a$ plays the role of a Lagrange multiplier enforcing the Gauss law constraints $\mathcal{C}_a$. The Hamiltonian density is:
\br
{\cal H}&=&\frac{1}{2}{\rm Tr}({\widetilde F}^2)+i{\bar\psi}\gamma_1\,D_1\psi+m{\bar\psi}\,\psi
+(\vp_1^{\dagger}\,D_1\vp + (D_1 \vp)^\dagger \vp_1)
\nonumber\\
&+&V\(\mid \vp\mid\)+2\pi_{\vp}\,\pi_{\vp^{\dagger}} - \vp_1^\dagger \vp_1.
\lab{hamiltonian}
\er
The Gauss law constraints $\mathcal{C}_a$ are given by
\be
{\cal C}_a=(D_1 {\widetilde F})^a-e\,J_0^a
\lab{sec4:constraint}
\ee
where $J_0^a$ is the time component of the non-abelian matter current:

\be
J_0 \equiv \bar{\psi}\,\gamma_0\,R_{\psi} -i\,[\pi_\varphi^\dagger\(R_\vp(T_a)\,\vp\) - \(\vp^\dagger\,R_{\vp}(T_a)\) \,\pi_\vp  ]\,T_a
\ee

In this formalism, the temporal components $A_0^a$ do not carry dynamics and instead enforce the constraints $\mathcal{C}_a = 0$, reflecting the local gauge invariance of the theory.

Variations of the action with respect to the auxiliary field $\widetilde{F}$ and the spatial gauge field $A_1$ lead to the equations of motion:
\br
{\widetilde F} &=& \partial_0 A_1 - \partial_1 A_0 + i\,e\,\sbr{A_0}{A_1}\\
D_0 {\widetilde F} &=&e\,J_1
\lab{sec4:eomfo}
\er
In addition, variations with respect to the auxiliary field $\vp_\m$ lead to the relation:
\be
\vp_\m = D_\m \vp.
\lab{vpauxeq}
\ee
Although we introduce $\vp_\m$ as auxiliary fields of the boson fields, in the context of the first-order formalism, the consistency with the gauge sector, where the time component of the gauge field plays the role of a Lagrangian multiplier, only requires that the time component of \rf{vpauxeq} is treated as a dynamical equation, The spatial components, i.e., $\vp_i = D_i \vp$, can be taken as an off-shell identity instead.

In such a formalism, the relation expressing the pseudoscalar $\widetilde{F}$ in terms of the gauge field does not follow directly from fundamental principles as a consequence of the gauge principle; rather, it emerges as a dynamical equation - an on-shell curvature relation. 

On the other hand, imposing the Gauss constraint, arising from the variation of the action with respect to $A_0$, gives
\be
{\cal C} = D_1  \widetilde{F} - e J_0 =0
\lab{sec4:constraintshold}
\ee
and the equations of motion of the first-order formalism \rf{sec4:eomfo} within the constraints equations \rf{sec4:constraintshold} describe the dynamics of the Yang-Mills theory \rf{sec2:em}.

From the Lagrangian density \rf{sec4:legt}, one obtains the total Hamiltonian including the primary constraints:
\be
H_T=\int_{-\infty}^{+\infty} dx\, \( {\cal H}-A_0^a\,{\cal C}_a\).
\lab{sec4:completeham}
\ee
The nonvanishing equal-time Poisson brackets between the canonical variables are:
\br
\{A_1^a(x)\,,\,\pi^b(y)\} &=& \{A_1^a(x)\,,\,{\widetilde F}^b(y)\}=\delta^{ab}\,\delta(x-y),\nonumber\\
\{\psi^{(f)}_{A}(x)\,,\,\pi^\psi_{(g),B}(y)\} &=& \delta_{fg}\,\delta_{AB}\,\delta(x-y),\nonumber\\
\{\vp(x)\,,\,\pi_\vp(y)\} &=& \delta(x-y),\nonumber\\
 \{\vp^\dagger(x)\,,\,\pi_\vp^\dagger(y)\} &=& \delta(x-y),
\lab{pbrel}
\er
where we have omitted the group indices of the representations $R^\psi$ and $R^\varphi$ for brevity; these are implicitly treated as Kronecker deltas on the right-hand side.

The matter currents form a representation of the gauge algebra, which can be verified using the brackets above:
\be
\{J_0^a(x)\,,\,J_0^b(y)\} =  f_{abc}\,J_0^c\,\delta(x-y).
\lab{sec4:gcopy}
\ee
By using the bracket relations \rf{pbrel} and \rf{curr}, one finds that the brackets give the transformations of the $J_0$ on the matter fields:
\br
\pbr{J_0(x)}{\psi(y)}&=& i\,[R_\psi(T_a)\,\psi(y)]\,\delta(x-y)\nonumber\\
\pbr{J_0(x)}{\pi_\psi(y)}&=& -i\,[\pi_\psi(y)\,R_\psi(T_a)]\,\delta(x-y)\nonumber\\
\pbr{J_0(x)}{\vp(y)}&=&i\,[R_\vp(T_a)\,\vp(y)]\,\delta(x-y) \nonumber\\
\pbr{J_0(x)}{\pi_\vp(y)}&=& -i\,[\pi_\vp(y)\,R_\vp(T_a)]\,\delta(x-y)
\lab{conm}
\er
From the same canonical structure, the constraints satisfy the following Poisson brackets:
\br
\{{\cal C}_a(x)\,,\,A_1^b(y)\} &=& -\delta_{ab}\,\frac{\partial}{\partial x} \delta(x-y) -e\, f_{abc}A_{1}^c(x)\,\delta(x-y)\nonumber\\
\{{\cal C}_a(x)\,,\,\widetilde{F}^b(y)\} &=& - e\, f_{abc}\,\widetilde{F}^c(x)\,\delta(x-y)\nonumber\\
\{{\cal C}_a(x)\,,\,J^b_0(y)\} &=&-e\,f_{abc}\,J_0^c(x) \delta(x-y)\nonumber\\
\{{\cal C}_a(x)\,,\,\psi(y)\} &=&- i\,e\,[R_\psi(T_a)\,\psi(y)]\,\delta(x-y)\nonumber\\
 \{{\cal C}_a(x)\,,\,\vp(y)\} &=&-i\,e\, [R_{\vp}(T_a)\,\vp(y)]\,\delta(x-y)
\lab{sec4:constrel}
\er
These relations show that the constraints generate infinitesimal gauge transformations, up to total derivatives. To make this explicit, consider the smeared constraint:
\be
{\cal G}[\alpha] \equiv \int_{-\infty}^{+\infty} dx\, \alpha^a\,C_a(x)
\lab{sec4:calg}
\ee
with $\alpha(x) = \alpha^a(x) \, T_a$ an element of the Lie algebra of $G$. The action of $\mathcal{G}[\alpha]$ on the matter fields gives:
\br
\{{\cal G}[\alpha]\,,\,J^b_0(y)\} &=&-e\,f_{abc}\,\alpha^a(y)\,J_0^c(y)\nonumber\\
\{{\cal G}[\alpha]\,,\,\psi(y)\} &=& -i\,e\,[R_\psi(\alpha(y))\,\psi(y)]\nonumber\\
 \{{\cal G}[\alpha]\,,\,\vp(y)\} &=&- i\,e\,[R_{\vp}(\alpha(y))\,\vp(y)],
\lab{sec4:calgmf}
\er 
which reproduces the expected infinitesimal gauge transformations under $G$, acting through the representations $R_\psi$ and $R_\varphi$. It corresponds to the copy of $G$ generated by the matter currents $J_0$ in \rf{sec4:gcopy}, since the contribution of the transformation under ${\cal C}_a$ comes from $J_0^a$ in \rf{sec4:constraint}, as the matter fields commute with $D_1 \widetilde{F}$. 
For the gauge fields, the action of $\mathcal{G}[\alpha]$ yields:
\br
\{{\cal G}[\alpha]\,,\,A^b_1(y)\} &=& (D_1 \alpha)^b(y) - \int_{-\infty}^{+\infty}dx\, \frac{\partial}{\partial x}[\alpha^a(x)\,\delta(x-y)]\nonumber\\
\{{\cal G}[\alpha]\,,\,\widetilde{F}^b(y)\} &=& -e\,f_{bca}\,\alpha^a(y)\,\widetilde{F}^c(y).
\lab{sec4:ggauget}
\er

The presence of a boundary term in the transformation of $A_1$ shows that $\mathcal{G}[\alpha]$ generates proper gauge transformations only if this term vanishes. This is ensured either by requiring the gauge parameter $\alpha^a(x)$ to vanish at spatial infinity (large gauge transformations), or by imposing periodic boundary conditions such that $\alpha(+\infty) = \alpha(-\infty)$.

\section{The symmetries generated by the conserved charges}
\label{sec:4}
\setcounter{equation}{0}

In gauge theories, symmetries play a central role in shaping the dynamics and structure of the physical system.  In (3+1)-dimensional Yang-Mills theories, it has been shown that non-abelian conserved charges obtained from the integral formulation in loop space generate a novel class of symmetries that do not correspond to gauge redundancies \cite{ymintg}. These symmetries act nontrivially on the physical degrees of freedom, yet leave the Hamiltonian invariant, and are thus physically meaningful. Motivated by this observation, we now investigate whether a similar structure emerges in two-dimensional Yang-Mills theory.

\subsection{The charge operator}
In section \ref{sec:2}, we obtained the conserved charges \rf{sec2:cc} from the holonomy \rf{opdef} along the real axis, that is, a particular solution obtained through the holonomy equation \rf{holeqq}. The holonomy equation \rf{holeqq} along a path traversing the real line at fixed time $t$, $\mathbb{R}_t \in {\cal M}$, starting from $x_R^{(t)}=(t,-\infty)$, can be written as follows
\be
\frac{d Q (x,-\infty)}{dx} - {\cal A}(\b,x) \,  Q(x,-\infty) = 0,   \qquad Q(-\infty, -\infty) = Q_R, 
\lab{holequ}
\ee
with 
\be
{\cal A}(\b,x) \equiv W^{-1}(x,-\infty)\,K_1\,W(x,-\infty).
\lab{cona}
\ee
Where the dependence upon $\b$ in \rf{cona} comes from $K_1$ the spatial component of $K_\mu$ defined in \rf{sec2:lj}, i.e.,
\be
K_1(\b,x) = i\,e^2\,\b\,\int_{0}^{1}d\l {\rm e}^{i\,e\,\b\,\l\,\wti{F}} J_0(x) {\rm e}^{-i\,e\,\b\,\l\,\wti{F}}
\lab{k1e}
\ee
and the Wilson line $W(x)$ a solution of the ordinary differential equation:
\be
\frac{dW}{dx}+i\,e\,A_1(x) W = 0, \qquad \qquad W(x_R) = W_R. 
\lab{sholeq}
\ee
which is \rf{holequ} when the path is traversing the real line at a fixed time $t$, starting from $x_R^{(t)}$ up to $x$ along the path on the real line.

Integrating \rf{holequ} up to $+\infty$ from the reference point $x_R^{(t)}$, we have the path-ordered exponential 
\be
Q(+\infty, -\infty) = Q_{x_R^{(t)}}(t,\b) = P_1e^{\int_{-\infty}^{+\infty}{\cal A}(\b,x)}Q_R,
\lab{opcha}
\ee
where the reference point is taken at $x_R^{(t)}=(t, -\infty)$. When the equations of motion hold, the equation \rf{sec2:lj} yields, and so the integral equations \rf{sec2:hintaux}. Thus, the eigenvalues of $Q_{x_R^{(t)}}(\b)$ will correspond to the dynamically conserved charges of the theory. In our approach to the symplectic formalism, the equations of motion \rf{sec2:em} are obtained when the constraints \rf{sec4:constraintshold} are imposed, and the Hamilton-Jacobi equations yield. Hence, the operator $Q_{(x,-\infty)}$ can be expressed in terms of $V(x)$ through the integral form of the dynamical equations, as in \rf{sec2:hintaux}, where we now denote by:
\be
Q(x,-\infty) \approx V(x) = e^{ie\b W^{-1}(x)\,\widetilde{F}(x) W(x)}
\lab{opcont}
\ee 
where we omitted the dependence of the time $t$, the symbol $\approx$ denotes that the constraints are imposed, $W(x)$ is the Wilson line obtained from \rf{sholeq}, $V_R$ stands for $V(x_R)$, and the reference point is taken at spatial infinity $x_R = (t,-\infty)$. In addition, in such a regime, that is, when the constraints hold, the integration constant $Q_R$ becomes:
\be
Q_R \approx V_R = e^{ie\b W^{-1}_R\,\widetilde{F}(x_R) W_R}
\lab{qr}
\ee

The Poisson bracket at equal time, of any phase space function $X$, with the charge operator $Q(\b)$ (where the dependence upon the reference point and the time $t$ was omitted, i.e., $Q(\b) = Q(t,\b)$, see \rf{opcha}), can be derived from \rf{holequ} (see \rf{qrelA}) and is given by
\be
\pbr{X}{Q(\b)} =Q(\b)\,\int_{-\infty}^{+\infty} dx\, Q^{-1}(x,-\infty)\pbr{X}{{\cal A}(\b;x)}\,Q(x,-\infty)
\lab{qrel}
\ee
where
\be
\pbr{X}{{\cal A}(\b;x)} = W^{-1}(x)\pbr{X}{K_1(x)}W(x)+\sbr{W^{-1}(x)K_1W(x)}{W^{-1}(x)\pbr{X}{W(x)}}
\lab{calax}
\ee
and
\br
\pbr{X}{K_1(x)} &=&i\,e^2\,\b \int_{0}^{1}d\l\,{\rm e}^{i\,e\,\b\,\l\,\wti{F}}\,\pbr{X}{J_0}\,{\rm e}^{-i\,e\,\b\,\l\,\wti{F}} \nonumber\\
&+&i\,e^2\,\b \int_{0}^{1}d\l\,{\rm e}^{i\,e\,\b\,\l\,\wti{F}}\sbr{{\rm e}^{-i\,e\,\b\,\l\,\wti{F}}\pbr{X}{{\rm e}^{i\,e\,\b\,\l\,\wti{F}}}}{J_0(x)}{\rm e}^{-i\,e\,\b\,\l\,\wti{F}}
\lab{k1rel}
\er
where the Poisson brackets $\pbr{X}{{\rm e}^{i\,e\,\b\,\l\,\wti{F}}}$ in the second term in the r.h.s. can be understood as a translation of $X$ on the canonical momentum direction in the phase space of the gauge fields, and can be evaluated through the expression\footnote{The derivation of \rf{pbl} follows the same reasoning of \rf{auxpbwx}, considering the equation $\frac{d}{d\l}{\rm e}^{i\,e\,\b\,\l\,\wti{F}} = ie\b\wti{F}\,{\rm e}^{i\,e\,\b\,\l\,\wti{F}}$}:
\be
\pbr{X}{{\rm e}^{i\,e\,\b\,\l\,\wti{F}}}= i\,e\,\b\,{\rm e}^{i\,e\,\b\,\l\,\wti{F}}\,\int_{0}^{\l}d\l^\prime\,{\rm e}^{-i\,e\,\b\,\l^\prime\,\wti{F}}\,\pbr{X}{\wti{F}}{\rm e}^{i\,e\,\b\,\l^\prime\,\wti{F}}.
\lab{pbl}
\ee

The Poisson bracket of $X$ with the Wilson line operator is given by (see \rf{pbwxf})
\be
W^{-1}(x)\,\{X\,,\,W(x)\} = -i\,e\,\lim_{x_R \rightarrow -\infty}\int_{x_R}^{x}dy\,\,W^{-1}(y)\,\{X\,,\,A_1(y)\}\,W(y)
\lab{wopxpb}
\ee
We compute the infinitesimal transformation generated by the non-abelian charges $q_N(\beta)$, given in \rf{sec2:cc}, on a phase space function $X$ by an equal time Poisson bracket 
\br
\delta_N X &\equiv& \vareps \{X\,,\,q_N(\beta)\}=i\,e\,\vareps\,{\rm Tr}\bigg[Q^N(\b) \int_{-\infty}^{+\infty}dx \,Q^{-1}(x,-\infty)\times \nonumber\\
&\times&\bigg(e\,\b \int_{0}^{1}d\l\,W^{-1}(x)\,{\rm e}^{i\,e\,\b\,\l\,\wti{F}}\,\pbr{X}{J_0}\,{\rm e}^{-i\,e\,\b\,\l\,\wti{F}}\,W(x) \nonumber\\
&+&i\,e^2\,\b^2 \int_{0}^{1}d\l\,\int_{0}^{\l}d\l^\prime \,W^{-1}(x)\,{\rm e}^{i\,e\,\b\,\l\,\wti{F}}\sbr{{\rm e}^{-i\,e\,\b\,\l^\prime\,\wti{F}}\pbr{X}{\wti{F}}{\rm e}^{i\,e\,\b\,\l^\prime\,\wti{F}}}{J_0(x)}{\rm e}^{-i\,e\,\b\,\l\,\wti{F}}W(x) \nonumber\\
&-&\sbr{W^{-1}(x)K_1W(x)}{\int_{-\infty}^x dy\,W^{-1}(y)\,\{X\,,\,A_1(y)\}\,W(y)}\bigg)Q(x,-\infty)\bigg]
\lab{tcc}
\er
with $\vareps$ being the infinitesimal parameter of the transformation.

The expression \rf{tcc} provides a general expression for evaluating the action of the charge on an arbitrary physical field $X$. It is useful for uncovering the underlying structure of these transformations. In particular, a similar pattern arises in the action of Yang–Mills charges in $3+1$ dimensions \cite{ymintg}, where the transformations receive contributions from both constraints and boundary terms. Although \rf{tcc} is not explicitly written in terms of the constraints, it offers a convenient framework for computing the Poisson brackets associated with the physical transformations, as discussed in the following sections.

Considering the equation \rf{sec4:constraint} to express the time component of the matter currents $J_0$ in terms of the constraint functional ${\cal C}$, i.e.,
\be
J_0 = \frac{1}{e}\(D_1 \wti{F} - {\cal C}\). 
\lab{jc}
\ee
it is then possible to rewrite \rf{tcc} in terms of ${\cal C}$, i.e., substituting \rf{jc} into \rf{tcc}, one finds that
\br
\delta_N X &\equiv& \vareps \{X\,,\,q_N(\beta)\}=i\,e\,\vareps\,{\rm Tr}\bigg[Q^N(\b) \int_{-\infty}^{+\infty}dx \,Q^{-1}(x,-\infty)\times \nonumber\\
&\times&\bigg(\b \int_{0}^{1}d\l\,W^{-1}(x)\,\,\pbr{X}{{\rm e}^{i\,e\,\b\,\l\,\wti{F}} D_1 \wti{F} {\rm e}^{-i\,e\,\b\,\l\,\wti{F}}}\,W(x)\nonumber\\
&-&\b \int_{0}^{1}d\l\,W^{-1}(x)\,{\rm e}^{i\,e\,\b\,\l\,\wti{F}}\,\pbr{X}{{\cal C}}\,{\rm e}^{-i\,e\,\b\,\l\,\wti{F}}\,W(x) \nonumber\\
&-&i\,e\,\b^2 \int_{0}^{1}d\l\,\int_{0}^{\l}d\l^\prime \,W^{-1}(x)\,{\rm e}^{i\,e\,\b\,\l\,\wti{F}}\sbr{{\rm e}^{-i\,e\,\b\,\l^\prime\,\wti{F}}\pbr{X}{\wti{F}}{\rm e}^{i\,e\,\b\,\l^\prime\,\wti{F}}}{{\cal C}(x)}{\rm e}^{-i\,e\,\b\,\l\,\wti{F}}W(x) \nonumber\\
&-&\sbr{W^{-1}(x)K_1W(x)}{\int_{-\infty}^x dyW^{-1}(y)\,\{X\,,\,A_1(y)\}\,W(y)}\bigg)Q(x,-\infty)\bigg]
\lab{tcca}
\er
Comparing the second line in \rf{tcca} with the integral representation of $L_\mu$ in \rf{sec2:lf}, we have that
\br
\delta_N X &\equiv& \vareps \{X\,,\,q_N(\beta)\}=-i\,e\,\vareps\,{\rm Tr}\bigg[Q^N(\b) \int_{-\infty}^{+\infty}dx \,Q^{-1}(x,-\infty)\times \nonumber\\
&\times&\bigg(\frac{i}{e} \pbr{X}{W^{-1}(x)L_1(x)W(x)}+\b \int_{0}^{1}d\l\,W^{-1}(x)\,{\rm e}^{i\,e\,\b\,\l\,\wti{F}}\,\pbr{X}{{\cal C}}\,{\rm e}^{-i\,e\,\b\,\l\,\wti{F}}\,W(x) \nonumber\\
&+&i\,e\,\b^2 \int_{0}^{1}d\l\,\int_{0}^{\l}d\l^\prime \,W^{-1}(x)\,{\rm e}^{i\,e\,\b\,\l\,\wti{F}}\sbr{{\rm e}^{-i\,e\,\b\,\l^\prime\,\wti{F}}\pbr{X}{\wti{F}}{\rm e}^{i\,e\,\b\,\l^\prime\,\wti{F}}}{{\cal C}(x)}{\rm e}^{-i\,e\,\b\,\l\,\wti{F}}W(x) \nonumber\\
&+&\sbr{W^{-1}(x)(K_1 - L_1)W(x)}{\int_{-\infty}^x dyW^{-1}(y)\,\{X\,,\,A_1(y)\}\,W(y)}\bigg)Q(x,-\infty)\bigg]
\lab{tcca1}
\er
Here we have used the properties of the Poisson brackets together with \rf{wopxpb} to absorb the Wilson line into the first bracket on the right-hand side. This bracket can be identified with the Poisson bracket of $X$ with the connection associated with $V(\b{x})$ through the holonomy equation \rf{sec2:v_eq}. Moreover, using the holonomy equation \rf{holequ}, one may perform an integration by parts on the first integral on the right-hand side, which leads to the following expression for \rf{tcca1}:
\br
&&\delta_N X \equiv \vareps \{X\,,\,q_N(\beta)\}=\nonumber\\
&&=\vareps {\rm Tr}\bigg[Q^{N}(\b)\bigg(\pbr{X}{V(\b,+\infty)}V^{-1}(\b,+\infty) - Q_R^{-1}\pbr{X}{V_R}V_R^{-1}Q_R\bigg)\bigg]\nonumber\\
&&-i\,e\,\vareps\,{\rm Tr}\bigg[Q^N(\b) \int_{-\infty}^{+\infty}dx \,Q^{-1}(x,-\infty)\bigg(\b \int_{0}^{1}d\l\,W^{-1}(x)\,{\rm e}^{i\,e\,\b\,\l\,\wti{F}}\,\pbr{X}{{\cal C}}\,{\rm e}^{-i\,e\,\b\,\l\,\wti{F}}\,W(x) \nonumber\\
&&+i\,e\,\b^2 \int_{0}^{1}d\l\,\int_{0}^{\l}d\l^\prime \,W^{-1}(x)\,{\rm e}^{i\,e\,\b\,\l\,\wti{F}}\sbr{{\rm e}^{-i\,e\,\b\,\l^\prime\,\wti{F}}\pbr{X}{\wti{F}}{\rm e}^{i\,e\,\b\,\l^\prime\,\wti{F}}}{{\cal C}(x)}{\rm e}^{-i\,e\,\b\,\l\,\wti{F}}W(x) \nonumber\\
&&+\sbr{W^{-1}(x)(K_1 - L_1)W(x)}{\frac{i}{e}\pbr{X}{V(x)}V^{-1}(x)+\int_{-\infty}^x dy W^{-1}(y)\pbr{x}{A_1(y)}W(y)}\bigg)\times\nonumber\\
&&\times Q(x,-\infty)\bigg]
\lab{tcca2}
\er
The pattern expressed in terms of the constraints ${\cal C}$, that is \rf{tcca2}, reveals in lower dimensions an underlying structure similar to that found in Yang-Mills in $3+1$ dimensions \cite{ymintg}, where boundary terms and constraints dictate the action generated by the charges.

\subsection{Transformation of matter fields}

To obtain the transformations of the matter fields generated by the conserved charges, we use that the matter fields commute, via Poisson brackets, with functionals of the gauge fields (see \rf{pbrel} and \rf{wopxpb}). Hence, using the relations \rf{conm} and considering the general expression \rf{tcc}, we obtain that the infinitesimal transformations on the matter fields generated from the conserved charges are the following
\be
\d_N \Psi(x) =\vareps\,e^2\,\b [R_{\Psi}(\xi_N^a(\b;x) T_a)\Psi(x)]
\lab{mft}
\ee
where we denoted $\Psi$ as being either the fermion fields ($\psi$) or the boson fields ($\vp$), and used \rf{conm} and defined
\be
\xi^a_N(\b;x) \equiv {\rm Tr}\bigg[Q^N\,Q^{-1}(x,-\infty)\,W^{-1}(x)\(\int_{0}^{1}d\l\,{\rm e}^{i\,e\,\b\,\l\,\wti{F}}T_a\,{\rm e}^{-i\,e\,\b\,\l\,\wti{F}}\) W(x)\,Q(x,-\infty)\bigg]
\lab{xiph}
\ee
Similarly, by using the relations \rf{conm}, one can find that
\be
\d_N \pi_\Psi = -\vareps\,e^2\,\b[\pi_\Psi\,R_{\Psi}(\xi_N^a(\b;x)\,T_a)]
\lab{mcct}
\ee
The phase element $\xi$ is a non-integrable phase since it depends upon Wilson lines and $Q$ operators along a path that joins the reference point to the point where the matter field is located. When the constraints are imposed, the weak relation \rf{opcont} yields, and so one can write
\be
\xi^N_a(\b;x) \approx   {\rm Tr}\bigg[Q^N \,W^{-1}(x)\(\int_{0}^{1}d\l\,{\rm e}^{i\,e\,\b\,(\l-1)\,\wti{F}}T_a\,{\rm e}^{-i\,e\,\b\,(\l-1)\,\wti{F}}\) W(x) \bigg]
\lab{xiphc}
\ee
with $W(x)$ obtained from \rf{sholeq}.

\subsection{Transformation of the gauge fields}
Consider the canonical brackets in \rf{pbrel} and the fact that the gauge field commutes with the matter currents via Poisson brackets, using the general expression \rf{tcc} for the transformations of the conserved charges, one gets the following
\br
&&\d_N A^a_1 = -\vareps\,e^3\,\b^2\,{\rm Tr}\bigg[Q^{N}(\b)Q^{-1}(x,-\infty)\,W^{-1}(x)\lab{acctr0}\times\\
&&\times\(\int_{0}^{1}d\l\,\int_{0}^{\l}d\l^\prime \,\sbr{{\rm e}^{i\,e\,\b\,(\l-\l^\prime)\,\wti{F}}T_a{\rm e}^{-i\,e\,\b\,(\l -\l^\prime)\,\wti{F}}}{{\rm e}^{i\,e\,\b\,\l\,\wti{F}}\,J_0(x)\,{\rm e}^{-i\,e\,\b\,\l\,\wti{F}}}\)W(x)\,Q(x,-\infty)\bigg]. \nonumber
\er
Using that
\be
\int_{0}^{\l}d\l^\prime {\rm e}^{i\,e\,\b\,(\l-\l^\prime)\,\wti{F}}T_a{\rm e}^{-i\,e\,\b\,(\l -\l^\prime)\,\wti{F}} = - \int_{0}^{\l}d\l^\prime {\rm e}^{i\,e\,\b\,\l^\prime\,\wti{F}}T_a{\rm e}^{-i\,e\,\b\, \l^\prime\,\wti{F}}
\ee
and by an integration by parts, one can rewrite the expression \rf{acctr0} as follows
\br
&&\d_N A^a_1 = \lab{acctr1}\\
&&=-i\,\vareps\, e\,\b\,{\rm Tr}\bigg[Q^{N}(\b)Q^{-1}(x,-\infty)W^{-1}(x)\bigg( \sbr{\int_{0}^{1}d\l\,{\rm e}^{i\,e\,\b\,\l\,\wti{F}}T_a{\rm e}^{-i\,e\,\b\,\l \,\wti{F}}}{K_1(x)} \nonumber\\
&&-i\,e^2\,\b\,\int_{0}^{1}d\l\int_{0}^{\l}d\l^\prime \,\sbr{{\rm e}^{i\,e\,\b\,\l\,\wti{F}}T_a{\rm e}^{-i\,e\,\b\,\l\,\wti{F}}}{{\rm e}^{i\,e\,\b\,\l^\prime\,\wti{F}}J_0(x){\rm e}^{-i\,e\,\b\,\l^\prime\,\wti{F}}}\bigg)W(x)Q(x,-\infty)\bigg]
\nonumber
\er
Note that, from the holonomy equations \rf{holequ} and \rf{sholeq} of $Q(x,-\infty)$ and $W(x)$, respectively, and using the definition of the constraint \rf{acctr1}, one can express the transformation \rf{acctr1} in terms of \rf{xiph}:
\br
&&\d_N A^a_1 =i\,\vareps\,e\,\b\, (D_1 \xi_N)^a - \vareps\,e^2\,\b^2\,{\rm Tr}\bigg[Q^{N}(\b)Q^{-1}(x,-\infty)\,W^{-1}(x)\times \lab{acctr}\\
&&\times \int_{0}^{1}d\l\,\int_{0}^{\l}d\l^\prime \,\sbr{{\rm e}^{i\,e\,\b\,\l\,\wti{F}}T_a{\rm e}^{-i\,e\,\b\,\l\,\wti{F}}}{{\rm e}^{i\,e\,\b\,\l^\prime\,\wti{F}}\,{\cal C}(x)\,{\rm e}^{-i\,e\,\b\,\l^\prime\,\wti{F}}}\bigg)W(x)\,Q(x,-\infty)\bigg]\nonumber
\er
where we used that
\br
&&\frac{d}{dx}\(W^{-1}{\rm e}^{i\,e\,\b\,\l\,\wti{F}}T_a{\rm e}^{-i\,e\,\b\,\l\,\wti{F}}W\) = 
i\,e\, W^{-1}{\rm e}^{i\,e\,\b\,\l\,\wti{F}}\sbr{A_1}{T_a} {\rm e}^{-i\,e\,\b\,\l\,\wti{F}}W\nonumber\\
&&+i\,e\,\b\,W^{-1}\sbr{\int_{0}^{\l}d\l^\prime{\rm e}^{i\,e\,\b\,\l^\prime\,\wti{F}}D_\m \wti{F}  {\rm e}^{-i\,e\,\b\,\l^\prime\,\wti{F}}}{{\rm e}^{i\,e\,\b\,\l\,\wti{F}} T_a {\rm e}^{-i\,e\,\b\,\l\,\wti{F}}}W.
\lab{dexp}
\er


The phase space flow of the Wilson operator, along the real line, is dictated by the flow of the gauge field component $A_1$ as shown in \rf{wopxpb}. Considering now, $X$ as an element of the Lie algebra, i.e., not as components of a matrix, but a matrix itself, the Poisson brackets will take values in the product of Lie algebras, $\mathfrak{g}\otimes \mathfrak{g}$. Hence, using \rf{wopxpb}, with $X = q_N(\b)$, and then considering the result \rf{acctr}, one gets that
\br
&&\d_N W(x) =\vareps\,e^2\,\b\( \xi_N(\b;x) \, W(x)  -W(x)\,\xi_N(\b;x_R) \)
\lab{wlct}\\
&&+i\, \vareps\,e^3\,\b^2\,W^{-1}(x)\int_{-\infty}^{x}dz\, W^{-1}(z)T_aW(z) \, {\rm Tr}\bigg[Q^{N}(\b)Q^{-1}(z,-\infty)\,W^{-1}(z) \times\nonumber\\
&&\times\int_{0}^{1}d\l\,\int_{0}^{\l}d\l^\prime \,\sbr{{\rm e}^{i\,e\,\b\,\l\,\wti{F}}T_a{\rm e}^{-i\,e\,\b\,\l\,\wti{F}}}{{\rm e}^{i\,e\,\b\,\l^\prime\,\wti{F}}\,{\cal C}(z)\,{\rm e}^{-i\,e\,\b\,\l^\prime\,\wti{F}}}\bigg)W(z)\,Q(z,-\infty)\bigg]
\nonumber
\er
where $x_R$ is the reference point taken at the spatial minus infinity at fixed time $t$, i.e., $(t,-\infty)$, and the element $\xi = \xi_a T_a$  at $x_R$ becomes 
\be
\xi^a_N(\b;x_R) \equiv {\rm Tr}\bigg[Q^N\,Q_R\,W^{-1}_R\(\int_{0}^{1}d\l\,{\rm e}^{i\,e\,\b\,\l\,\wti{F}(x_R)}T_a\,{\rm e}^{-i\,e\,\b\,\l\,\wti{F}(x_R)}\) W_R\,Q_R\bigg]
\lab{xiref}
\ee

In order to evaluate the full action of the charges on the phase space of the gauge fields, we need to consider the action on the conjugate momenta of the gauge field $A_1(x)$, that is, the pseudoscalar field ${\widetilde F}$. From \rf{tcc}, only the brackets with $A_1$ contribute for a non-vanishing transformation of $\wti{F}$ under the charges via Poisson brackets, resulting in what follows 
\be
\d_N \wti{F}^a = i\,e\,\vareps {\Tr}\(Q^N(\b)\int_{x}^{+\infty} dz \, Q^{-1}(z,-\infty)\sbr{W^{-1}(z)K_1W(z)}{W^{-1}(x)\,T_a\,W(x)}Q(z,-\infty)\)
\ee
where we used that
\be
\int_{-\infty}^{z}dy \,\d(x-y) = \th(z-x)
\ee
with $\th(z-x)$ the step function.
Using \rf{holequ}, we obtain that the transformation of the field ${\widetilde F}(x)$ generated by the charges is given by
\be
\delta_N {\widetilde F}^a(x) =ie\vareps{\rm Tr}\bigg[Q^N(\b)\bigg( Q^{-1}(x,-\infty)W^{-1}(x)T_aW(x)Q(x,-\infty)-W^{-1}(x)T_aW(x)\bigg)\bigg]
\lab{stfttrans1}
\ee

\subsection{On-shell transformations}
When the constraints \rf{sec4:constraintshold} are imposed, the transformations of the gauge field variables under the charges become simple and can be expressed as a gauge transformation by the phase factor $\xi_N$. For the gauge field $A_1$ and the Wilson line $W(x)$, one can notice it directly from \rf{acctr} and \rf{wlct} when ${\cal C}=0$, i.e.,
\br
\d_N A^a_1(x) &\approx& i \,\vareps\,e\,\b \(D_1 \xi\)^a \nonumber\\
\d_N W(x) &\approx&  \vareps\,e^2\,\b\( \xi_N(\b;x) \, W(x)  -W(x)\,\xi_N(\b;x_R) \)
\lab{gfxi}
\er
and the transformations of the matter fields remain the same, i.e., given by \rf{mft}.

For $\wti{F}$, when \rf{sec4:constraintshold} holds true, that is, the constraints are imposed, the integration constant $Q_R$ can be expressed as $V_R$, see \rf{qr}, and so, one can rewrite \rf{stfttrans1} as follows
\br
\d_N \wti{F}^a &\approx& -i\,\vareps\,e {\rm Tr}\(Q^N(\b)W^{-1}(x)\int_{0}^1 d\l\,\frac{d}{d\l}\({\rm e}^{i\,e\,\b\,(\l-1)\,\wti{F}}T_a{\rm e}^{-i\,e\,\b\,(\l-1)\,\wti{F}}\)W(x) \) \nonumber\\
&=& \vareps\,e^2\,\b {\rm Tr}\(Q^N(\b)\,V(\b,x)\,\int_{0}^1 d\l\,{\rm e}^{i\,e\,\b\,\l\,\wti{F}}\sbr{\wti{F}}{T_a}{\rm e}^{-i\,e\,\b\,\l\,\wti{F}}V(\b,x) \)
\er
where we used that $V(\b,x) = {\rm e}^{i\,e\,\b \wti{F}(x)}$. Hence, we obtain that 
\be
\d_N \wti{F}^a \approx -i\,\vareps\,e^2\,\b\,f_{abc} \wti{F}^b\,\xi_N^c(\b;x)
\lab{stfttrans2}
\ee

Although the charge transformations, on-shell, share a similar structure with the gauge transformations, they are not the same. Unlike the gauge transformations, charge transformations leave the vacuum configurations invariant.

The coupling of the Wilson line to the physical fields promotes the local gauge symmetry to a global symmetry, where the local character of the gauge transformations is absorbed into the Wilson line, and the gauge transformations act on a fixed reference point $x_R$ (see \rf{sec2:wgt}). In fact, this coupling is nothing less than the parallel transport by the Wilson line, and it acts on the physical fields depending upon the representation of the gauge group $G$ acting on those fields. For matter fields and the strength tensor, the parallel transport by the Wilson line is then given by
\be
\Psi^W \equiv R_\Psi(W^{-1})\,\Psi; \qquad \qquad F^W_{\m \n} \equiv  W^{-1}\,F_{\m \n}\, W 
\ee
Hence, under gauge transformations, using \rf{sec2:wgt}, we have that
\be
\Psi^W \rightarrow g(x_R)\,\Psi^W, \qquad \qquad F^W_{\m \n} \rightarrow g(x_R)\,F^W_{\m \n}\,g^{-1}(x_R).
\ee

The transport by the Wilson line, which is a solution of \rf{sholeq}, also removes the non-integrable character of the transformations obtained in previous sections. Thus, considering the transformations \rf{mft}, \rf{wlct} and \rf{stfttrans2} we obtain that
\br
\d_N \Psi^W \approx  [R_\psi(\xi(\b;x_R)) \Psi^W] \nonumber\\
\d_N {\widetilde F}^W \approx   \sbr{\xi(\b;x_R)}{{\widetilde F}^W} 
\er
where we used that $\xi(\b;x_R)= \xi_a(\b;x_R)\,T_a$, with $\xi_a(\b;x_R)$ give in \rf{xiref}.

\subsection{Symmetry of the Hamiltonian}
The conserved charges introduced in the previous sections arise from the geometric structure of the integral Yang-Mills equation in loop space. In particular, they are based on the requirement of path independence of the holonomy, a condition analogous to the zero-curvature condition in integrable field theories. While their construction is manifestly covariant and non-local, we now examine their role within the canonical formalism. To determine whether these charges correspond to genuine symmetries of the theory, we must verify that they generate transformations compatible with the system's Hamiltonian dynamics. In the Hamiltonian framework, this requires that the total Hamiltonian $H_T$ be invariant under the transformations generated by $q_N(\beta)$, up to terms vanishing on the constraint surface:
\be
\delta_N H_T = \vareps\, \{H_T,\, q_N(\beta)\} \approx 0.
\lab{sec5:hts}
\ee
This condition ensures that the action of the charges preserves the evolution of the physical degrees of freedom. Crucially, we do not demand strict invariance of $H_T$; invariance on the constraint surface is sufficient, as the true dynamics of the theory unfold only after the first-class constraints are imposed (cf.~discussion below~\rf{sec4:constraintshold}).

Decomposing the total Hamiltonian as
\be
H_T = H_G + H_{\psi} + H_{\vp} - H_C,
\ee
with
\br
 H_G &=& \frac{1}{2}\, \int_{-\infty}^{+\infty}dx\,{\rm Tr}\({\widetilde F}^2\),\qquad \qquad H_{C} = \int_{-\infty}^{+\infty}dx\,A_0^a \,C_a, \nonumber\\
 H_{\psi} &=& \int_{-\infty}^{+\infty}dx\,\(i\, \bar{\psi}\,\gamma_1 D_1 \psi + m\,\bar{\psi}\,\psi\)\nonumber\\
 H_{\vp} &=& \int_{-\infty}^{+\infty}dx\,\(\(D_1\vp\)^{\dagger}\,D_1\vp + V\(|\vp|\) + 2\,\pi_\vp\,\pi_\vp^\dagger \),
\lab{sec5:sectorh}
\er
we now proceed to show that the invariance condition \rf{sec5:hts} holds not only for the total Hamiltonian but also extends to each of its constituent sectors, where $H_G$ denotes the pure gauge sector, $H_{\psi}$ and $H_{\vp}$ correspond to the fermionic and scalar matter sectors respectively, and $H_C$ encodes the constraints. 

\subsubsection{Matter sector}
We find that the matter fields transform by a phase factor under the conserved charges \rf{mft} and \rf{mcct}, as a result, any composed state by the physical fields $\bar{\psi}$, $\psi$, $\vp$,$\vp^\dagger$, $\pi_\vp$, and $\pi_\vp^\dagger$ that are gauge-invariant, is also invariant under the transformations generated by the charges, so we get that
\be
\d_N (m \, \bar{\psi}\,\psi) =0, \qquad \qquad \d_N V(|\vp|) = \frac{\d V}{\d |\vp|}\,\d_N |\vp| =0, \qquad \qquad \d_N(\pi_\vp \,\pi_\vp^\dagger) = 0 
\lab{mfieldss}
\ee

Moreover, the transformation of the kinetic terms  of $H_\vp$ and $H_\psi$ under the charges depends on the transformations of
$D_1 \phi$ and $D_1 \vp$ under the charges. Denoting by $\Psi$  both fermionic ($\psi$) and bosonic fields ($\vp$), respectively, $R_\Psi$, their representations under the gauge group action, and using the transformation \rf{mft}, we have the following
\br
\d_N\(D_1 \Psi\) &=& \partial_1 (\d_N\,\Psi) + ieR_\Psi\,(\d_N A_1)\Psi + ieR_\Psi(A_1)\,\d_N\Psi\nonumber\\
&=&[R_\Psi(\xi)D_1 \Psi] + [R_\Psi\(D_1 \xi + ie\d_N A_1 \)\Psi]
\lab{covd}
\er
where $\xi = \xi_a T_a$ is given in \rf{xiph} and $D_1 = \partial_1 + i\,e\, R_\Psi(A_1)$. Thus, using \rf{gfxi} into \rf{covd}, we get that the covariant derivative of the matter fields transforms as
\be
\delta_N (D_1 \Psi) \approx [R_\Psi(\xi)\,D_1 \Psi], \qquad \qquad \Psi = \psi, \vp
\lab{covmt}
\ee

Consequently, from \rf{mft} and \rf{covmt}, the kinetic terms built from covariant derivatives are invariant:
\be
\delta_N\(\bar{\psi}\gamma_1 D_1\psi\) \approx \delta_N\(\(D_1\vp\)^\dagger D_1\vp\) \approx 0
\lab{sec5:kine}
\ee

Combining \rf{mfieldss} and \rf{sec5:kine}, we conclude that the matter sector of the Hamiltonian is invariant under the transformations generated by the conserved charges:
\be
\delta_N H_\psi \approx \delta_N H_\vp \approx 0
\lab{sec5:hm}
\ee
This result shows that the hidden symmetries encoded by $q_N(\beta)$ preserve the dynamics of both the fermionic and scalar matter fields.

\subsubsection{Gauge sector}
The gauge sector of the total Hamiltonian is given by $H_G$ as defined in \rf{sec5:sectorh}. Its variation under the transformations generated by the conserved charges reads
\be
\delta_N H_G = \int_{-\infty}^{+\infty}dx\, {\rm Tr}({\widetilde F}\,\delta_N {\widetilde F}) = \int_{-\infty}^{+\infty}dx\,{\widetilde F}^a\,\delta_N{\widetilde F}^a
\lab{sec5:deltahg}
\ee
Since the transformation $\delta_N {\widetilde F}^a$, on-shell, is under the adjoint action of the gauge group, as given in \rf{stfttrans2}, we conclude that
\be
\delta_N H_G \approx 0.
\lab{sec5:hg}
\ee
Thus, under the imposition of the constraints, the gauge sector of the Hamiltonian remains invariant under the transformations generated by the non-abelian conserved charges.

\subsubsection{Constraints sector}
To evaluate the transformation of the constraint sector of the total Hamiltonian  ($H_C$), given in \rf{sec5:sectorh}, we consider the pattern in \rf{qrel}, such that, using \rf{calax}, we have to evaluate the Poisson brackets
\br
\pbr{H_C}{{\cal A}(\b;x)} &=& W^{-1}(x)\pbr{H_C}{K_1(x)}W(x)\nonumber\\
&+&\sbr{W^{-1}(x)K_1W(x)}{W^{-1}(x)\pbr{H_C}{W(x)}}
\lab{hcal}
\er
Notice, following the definition \rf{sec4:calg}, by taking $\a = A_0$, $H_C$ can be written as the operator ${\cal G}[A_0]$ such that for any functional $X$ of the physical fields, we have that
\be
\pbr{H_C}{X} = \pbr{{\cal G}[A_0]}{X}.
\lab{hgx}
\ee

Hence, from \rf{sec4:calgmf}, we have that
\br
\{{\cal G}[A_0]\,,\,J_0^b(y)\} &=& - e\,f_{abc}\,A_0^a(y)\,J_0^c(y) \nonumber\\
\pbr{{\cal G}[A_0]}{{\widetilde F}^b(y)} &=& - e\,f_{abc}\,A_0^a(y)\,{\widetilde F}^c(y) 
\er
which can be rewritten as follows
\br
\{{\cal G}[A_0]\,,\,J_0^b(y)\}T_b &=&- i\,e\,\sbr{A_0(y)}{J_0(y)}\nonumber\\
\pbr{{\cal G}[A_0]}{{\widetilde F}^b(y)}T_b &=& - i\,e\,\sbr{A_0(y)}{{\widetilde F}(y)}
\lab{hcj}
\er
Considering the general Poisson brackets relation with $K_1$  given in \rf{k1rel}, and by using \rf{hcj}, we get that
\br
&&\pbr{{\cal G}[A_0]}{K_1(y)}=e^3\,\b \int_{0}^{1}d\l\,{\rm e}^{i\,e\,\b\,\l\,\wti{F}}\,\sbr{A_0(y)}{J_0(y)}\,{\rm e}^{-i\,e\,\b\,\l\,\wti{F}} \lab{GK}\\
&&+i\,e^4\,\b^2 \int_{0}^{1}d\l\int_{0}^\l d\l^\prime\,{\rm e}^{i\,e\,\b\,\l\,\wti{F}}\sbr{{\rm e}^{-i\,e\,\b\,\l^\prime\,\wti{F}}\sbr{A_0(y)}{{\widetilde F}(y)}{\rm e}^{i\,e\,\b\,\l^\prime\,\wti{F}}}{J_0(x)}{\rm e}^{-i\,e\,\b\,\l\,\wti{F}}.
\nonumber\er
Notice that one can write the following
\br
{\rm e}^{-i\,e\,\b\,\l\,\wti{F}}\sbr{A_0}{{\rm e}^{i\,e\,\b\,\l\,\wti{F}}} &=&\int_{0}^\l d\l^\prime \frac{d}{d\l}\({\rm e}^{-i\,e\,\b\,\l^\prime\,\wti{F}}A_0 {\rm e}^{i\,e\,\b\,\l^\prime\,\wti{F}}\)\nonumber\\
&=&-\frac{i}{e\,\b}\int_{0}^\l d\l^\prime {\rm e}^{-i\,e\,\b\,\l^\prime\,\wti{F}}\sbr{A_0(y)}{{\widetilde F}(y)}{\rm e}^{i\,e\,\b\,\l^\prime\,\wti{F}}
\lab{auxp}
\er
Thus, using the properties of Lie commutators, \rf{k1e}, and \rf{auxp}, \rf{GK} can be rewritten as follows
\be
\pbr{{\cal G}[A_0]}{K_1(y)}= ie\sbr{A_0(y)}{K_1(y)}
\lab{hk12}
\ee
Following the expression of Poisson brackets with the Wilson operator given in \rf{wopxpb}, we can evaluate the Poisson brackets: 
\be
W^{-1}(y)\{{\cal G}[A_0]\,,\,W(y)\} = -i\,e\int_{-\infty}^{y}dz\,W^{-1}(z)\,T_a\,W(z)\,\{{\cal G}[A_0]\,,\,A_1^a(z)\}
\lab{sec5:gw}
\ee
By using the transformation of $A_1$ given in \rf{sec4:ggauget} into \rf{sec5:gw}, we have that
\br
W^{-1}(y)\{{\cal G}[A_0]\,,\,W(y)\} &=& -i\,e\left[W^{-1}(y)\,A_0(y)\,W(y) - A_0(-\infty)\right.\\
&-&\left. \int_{-\infty}^{+\infty}dx\,\frac{\partial}{\partial x}\(W^{-1}(x)A_0(x)W(x)\,\theta(y-x)\)
\right]\nonumber
\er
where we used that $W(\infty) = \one$. Finally, performing the integral above, we get that
\be
W^{-1}(y)\{{\cal G}[A_0]\,,\,W(y)\} =-i\,e\,W^{-1}(y)\,A_0(y)\,W(y)
\lab{hcw}
\ee
By considering \rf{hgx}, substituting \rf{hk12} and \rf{hcw} into \rf{hcal}, we get that
\be
\pbr{H_C}{{\cal A}(\b;x)} = 0
\lab{hcalf}
\ee
Since the Poisson brackets \rf{hcalf} vanish,  we conclude, by using the pattern \rf{qrel}, and so \rf{tcc}, that the constraint sector is indeed invariant under the transformations generated by the charges, i.e.,
\be
\delta_N H_C = 0.
\lab{sec5:hc}
\ee

Therefore, from \rf{sec5:hm}, \rf{sec5:hg}, and \rf{sec5:hc}, we verify that the transformations generated by the charges satisfy \rf{sec5:hts} when the constraints \rf{sec4:constraint} are imposed.

\section{The Poisson algebra of the charges}
\label{sec:5}
\setcounter{equation}{0}
In higher-dimensional Yang-Mills models, the conserved charges commute via Poisson brackets and their associated charge operators satisfy a Sklyanin-like relation \cite{ymintg}, revealing structures that resemble integrability. 

Our definition of $Q$ obtained from \rf{holequ}, for an off-shell charge operator, relies on a non-local connection, whether in the loop space ${\cal L}^{0}$ or in spacetime. Such a condition does not lead to a proper Maillet bracket structure \cite{Maillet:1985ek, Maillet:1985fn} -- a general form for the Fundamental Poisson-Lie bracket (FPR) -- for the algebra of connections. Such a structure guarantees the involution of the charges. Still, we found that the charges Poisson commute under suitable conditions by investigating the algebra of the charge operators. A different definition of an off-shell charge operator can be taken, associated with a local connection; however, it does not follow the property of being covariant under gauge transformations, and, in fact, may not have regular gauge transformations off-shell.

Employing the pattern for the charge operator given in \rf{qrel}, the Poisson brackets, at equal time, for two charge operators associated with different parameters $Q(\b_1)$ and $Q(\b_2)$ are given by
\br
&&\pbro{Q(\b_1)}{Q(\b_2)}=Q(\b_1)\otimes Q(\b_2)\times \lab{qqpb}\\
&&\times\int_{-\infty}^{+\infty} dx \int_{-\infty}^{+\infty} dy\, Q_1^{-1}(x)\otimes Q_2^{-1}(y)\pbro{{\cal A}(\b_1;x)}{{\cal A}(\b_2;y)}\,Q_1(x) \otimes Q_2(y)
\nonumber
\er
where we denoted $Q_s(x)$, the holonomy $Q(x,-\infty)$ that depends upon the parameter $\b_s$, with $s=1,2$. By using \rf{calax}, one can write the bracket involving the connections ${\cal A}$, as follows
\br
\pbro{{\cal A}(\b_1;x)}{{\cal A}(\b_2;y)}&=&W^{-1}(x)\otimes W^{-1}(y)\pbro{K_1(\b_1;x)}{K_1(\b_2;y)}W(x)\otimes W(y)\nonumber\\
&+&{\cal I}_L + {\cal I}_R
\lab{aapb}
\er
with
\br
{\cal I}_L&=&\one\otimes W^{-1}(y)\times \lab{irle}\\
&&\times\sbr{W^{-1}(x)K_1(\b_1;x)W(x)\otimes \one}{W^{-1}(x)\otimes\one \pbro{W(x)}{K_1(\b_2;y)}}\one\otimes W(y)\nonumber\\
{\cal I}_R&=&W^{-1}(x)\otimes \one \times \nonumber\\
&&\times\sbr{\one \otimes W^{-1}(y)K_1(\b_2;y)W(y)}{\one \otimes W^{-1}(y) \pbro{K_1(\b_1;x)}{W(y)}} W(x)\otimes\one \nonumber
\er
where we used the fact that the Wilson lines commute with themselves via Poisson brackets. 

The first brackets in the right-hand side of \rf{aapb}, involving $K_1(\b_1)$ and $K_1(\b_2)$, can be evaluated using \rf{pbrel} and by considering the definition of $K_1$. The only non-vanishing contribution comes from brackets that involve the matter current terms, in the form of \rf{sec4:gcopy}, so one gets the following
\br
&&\pbro{K_1(\b_1;x)}{K_1(\b_2;y)} = -e^4\b_1\b_2 f_{cab} J_0^c\,\d(x-y)\lab{k1k2rel1}\\
&&\times\int_{0}^{1}d\l_1\int_{0}^{1}d\l_{2}\, {\rm e}^{i\,e\,\b\,\l_1\,\wti{F}(x)}\otimes {\rm e}^{i\,e\,\b\,\l_2\,\wti{F}(y)}\, T_a \otimes T_b \, {\rm e}^{-i\,e\,\b\,\l_1\,\wti{F}(x)}\otimes {\rm e}^{-i\,e\,\b\,\l_2\,\wti{F}(y)}.  
\nonumber
\er
which can be rewritten as follows
\br
&&\pbro{K_1(\b_1;x)}{K_1(\b_2;y)}=\d(x-y)\frac{e^2\b_1\b_2}{\b_1 - \b_2}\int_{0}^{1} d\l_1\int_{0}^{1}d\l_{2}\times \\
&&\times {\rm e}^{i\,e\,\b_1\,\l_1\,\wti{F}}\otimes {\rm e}^{i\,e\,\b_2\,\l_2\,\wti{F}}\, \sbr{T_a \otimes T_a}{ie^2\b_1\,J_0 \otimes \one + \one \otimes ie^2\b_2\,J_0} \, {\rm e}^{-i\,e\,\b_1\,\l_1\,\wti{F}}\otimes {\rm e}^{-i\,e\,\b_2\,\l_2\,\wti{F}}. 
\lab{k1k2rel2}\nonumber
\er
The evaluation of \rf{irle} is obtained using the general expression of the action via Poisson brackets of $K_1$, given in \rf{k1rel}, where it can be seen that only the Poisson brackets with ${\rm e}^{ie\b\l\wti{F}}$ survive in the evaluation, since the brackets are taken with the Wilson line, which commutes with the matter fields. Hence, one can write
\br
&&{\cal I}_L=i\,e^2\,\b_2\,\one\otimes W^{-1}(y)\bigg[W^{-1}(x)K_1(\b_1;x)W(x)\otimes \one\,,\,W^{-1}(x)\otimes\one  \times \nonumber\\
&&\times\int_{0}^{1}d\l \,\one \otimes {\rm e}^{i\,e\,\b_2\,\l\,\wti{F}(y)}\sbr{\one \otimes {\rm e}^{-i\,e\,\b_2\,\l\,\wti{F}(y)}\pbro{W(x)}{{\rm e}^{i\,e\,\b_2\,\l\,\wti{F}(y)}}}{\one\otimes J_0(y)}\one \otimes{\rm e}^{-i\,e\,\b_2\,\l\,\wti{F}(y)}\bigg]\times \nonumber\\
&&\times \one\otimes W(y)\nonumber\\
\lab{irle2}\\
&&{\cal I}_R=i\,e^2\,\b_1\,W^{-1}(x)\otimes \one \bigg[\one \otimes W^{-1}(y)K_1(\b_2;y)W(y)\,,\,\one \otimes W^{-1}(y)\times \nonumber\\
&&\times \int_{0}^{1}d\l {\rm e}^{i\,e\,\b_1\,\l\,\wti{F}(x)}\otimes \one \sbr{{\rm e}^{-i\,e\,\b_1\,\l\,\wti{F}(x)}\otimes \one \pbro{{\rm e}^{i\,e\,\b_1\,\l\,\wti{F}(x)}}{W(y)}}{J_0(x)\otimes \one} {\rm e}^{-i\,e\,\b_1\,\l\,\wti{F}(x)}\otimes \one\bigg]\times \nonumber\\
&&\times W(x)\otimes\one \nonumber
\er
Using the Poisson brackets expressions \rf{pbl} and \rf{wopxpb} to evaluate the brackets with the Wilson line in \rf{irle2}, we get that
\br
&&{\cal I}_L=i\,e^4\,\b_2^2\,\th(x-y)\,\sbr{W^{-1}(x)K_1(\b_1;x)W(x)}{W^{-1}(y)T_a W(y)}\otimes \nonumber\\
&&\otimes W^{-1}(y)\int_{0}^{1}d\l\int_{0}^{\l}d\l^\prime \, \sbr{{\rm e}^{i\,e\,\b_2\,(\l-\l^\prime)\,\wti{F}(y)}T_a{\rm e}^{-i\,e\,\b_2\,(\l-\l^\prime)\,\wti{F}(y)}}{{\rm e}^{i\,e\,\b_2\,\l\,\wti{F}(y)} J_0(y){\rm e}^{-i\,e\,\b_2\,\l\,\wti{F}(y)}} W(y)\nonumber\\
\lab{irle21}\\
&&{\cal I}_R=-i\,e^4\,\b_1^2\,\th(y-x)W^{-1}(x)\times \nonumber\\
&&\times\int_{0}^{1}d\l\int_{0}^{\l}d\l^\prime \,\sbr{ {\rm e}^{i\,e\,\b_1\,(\l-\l^\prime)\,\wti{F}(x)}T_a{\rm e}^{-i\,e\,\b_1\,(\l-\l^\prime)\,\wti{F}(x)}}{{\rm e}^{i\,e\,\b_1\,\l\,\wti{F}(x)} J_0(x){\rm e}^{-i\,e\,\b_1\,\l\,\wti{F}(x)}} W(x) \otimes \nonumber\\
&&\otimes \sbr{W^{-1}(y)K_1(\b_2;x)W(y)}{W^{-1}(x)T_a W(x)}\nonumber
\er
In addition, integrating by parts, we have that
\br
&&i\,e^2\,\b\,\int_{0}^{1}d\l\int_{0}^{\l}d\l^\prime \,\sbr{ {\rm e}^{i\,e\,\b\,(\l-\l^\prime)\,\wti{F}(x)}T_a{\rm e}^{-i\,e\,\b\,(\l-\l^\prime)\,\wti{F}(x)}}{{\rm e}^{i\,e\,\b\,\l\,\wti{F}(x)} J_0(x){\rm e}^{-i\,e\,\b\,\l\,\wti{F}(x)}} =
\nonumber\\
&&=\sbr{\int_{0}^{1}d\l^\prime \,{\rm e}^{i\,e\,\b\,\l^\prime\,\wti{F}(x)} T_a {\rm e}^{-i\,e\,\b\,\l^\prime\,\wti{F}(x)}}{K_1(\b;x)} \nonumber\\
&&- i\,e^2\,\b\,\int_{0}^{1}d\l^\prime\int_{0}^{\l^\prime}d\l\sbr{ {\rm e}^{i\,e\,\b\,\l^\prime\,\wti{F}(x)}T_a{\rm e}^{-i\,e\,\b\,\l^\prime\,\wti{F}(x)}}{{\rm e}^{i\,e\,\b\,\l\,\wti{F}(x)} J_0(x){\rm e}^{-i\,e\,\b\,\l\,\wti{F}(x)}}
\lab{intauxp}
\er
then, using \rf{intauxp} into \rf{irle2}, one gets that
\br
&&{\cal I}_L=e^2\,\b_2\,\th(x-y)\,\sbr{W^{-1}(x)K_1(\b_1;x)W(x)}{W^{-1}(y)T_a W(y)}\otimes \nonumber\\
&&\otimes W^{-1}(y)\bigg(\sbr{\int_{0}^{1}d\l^\prime \,{\rm e}^{i\,e\,\b_2\,\l^\prime\,\wti{F}(y)} T_a {\rm e}^{-i\,e\,\b_2\,\l^\prime\,\wti{F}(y)}}{K_1(\b_2,y)}\nonumber\\
&&-i\,e^2\,\b_2\,\int_{0}^{1}d\l^\prime\int_{0}^{\l^\prime}d\l\sbr{ {\rm e}^{i\,e\,\b_2\,\l^\prime\,\wti{F}(y)}T_a{\rm e}^{-i\,e\,\b_2\,\l^\prime\,\wti{F}(y)}}{{\rm e}^{i\,e\,\b_2\,\l\,\wti{F}(y)} J_0(y){\rm e}^{-i\,e\,\b_2\,\l\,\wti{F}(y)}}\bigg)W(y)\nonumber\\
\lab{irle3}\\
&&{\cal I}_R=-e^2\,\b_1\,\th(y-x)W^{-1}(x)\bigg(\sbr{\int_{0}^{1}d\l^\prime \,{\rm e}^{i\,e\,\b_1\,\l^\prime\,\wti{F}(x)} T_a {\rm e}^{-i\,e\,\b_1\,\l^\prime\,\wti{F}(x)}}{K_1(\b_1;x)} \nonumber\\
&&-i\,e^2\,\b_1\int_{0}^{1}d\l^\prime\int_{0}^{\l^\prime}d\l\sbr{ {\rm e}^{i\,e\,\b_1\,\l^\prime\,\wti{F}(x)}T_a{\rm e}^{-i\,e\,\b_1\,\l^\prime\,\wti{F}(x)}}{{\rm e}^{i\,e\,\b_1\,\l\,\wti{F}(x)} J_0(x){\rm e}^{-i\,e\,\b_1\,\l\,\wti{F}(x)}}\bigg)W(x)\otimes \nonumber\\
&&\otimes \sbr{W^{-1}(y)K_1(\b_2;y)W(y)}{W^{-1}(x)T_a W(x)}\nonumber
\er
In addition, writing $J_0$ in terms of the constraints \rf{sec4:constraint}, the expressions of ${\cal I}_{R/L}$ in \rf{irle3} can be written, separating the non-vanishing contribution when the constraints hold, i.e. ${\cal C}=0$, as follows
\br
&&{\cal I}_L=e^2\,\b_2\,\th(x-y)\,\sbr{W^{-1}(x)K_1(\b_1;x)W(x)}{W^{-1}(y)T_a W(y)}\otimes \nonumber\\
&&\otimes W^{-1}(y)\bigg(\sbr{\int_{0}^{1}d\l^\prime \,{\rm e}^{i\,e\,\b_2\,\l^\prime\,\wti{F}(y)} T_a {\rm e}^{-i\,e\,\b_2\,\l^\prime\,\wti{F}(y)}}{K_1(\b_2,y)}\nonumber\\
&&+\int_{0}^{1}d\l^\prime \, D_1\( {\rm e}^{i\,e\,\b_2\,\l^\prime\,\wti{F}(y)}T_a{\rm e}^{-i\,e\,\b_2\,\l^\prime\,\wti{F}(y)}\)-ie\int_{0}^{1}d\l^\prime \,  {\rm e}^{i\,e\,\b_2\,\l^\prime\,\wti{F}(y)}\sbr{A_1}{T_a}{\rm e}^{-i\,e\,\b_2\,\l^\prime\,\wti{F}(y)}\nonumber\\
&&-i\,e\,\b_2\,\int_{0}^{1}d\l^\prime\int_{0}^{\l^\prime}d\l\sbr{ {\rm e}^{i\,e\,\b_2\,\l^\prime\,\wti{F}(y)}T_a{\rm e}^{-i\,e\,\b_2\,\l^\prime\,\wti{F}(y)}}{{\rm e}^{i\,e\,\b_2\,\l\,\wti{F}(y)} {\cal C}(y){\rm e}^{-i\,e\,\b_2\,\l\,\wti{F}(y)}}\bigg)W(y)\nonumber\\
\lab{irle4}\\
&&{\cal I}_R=-e^2\,\b_1\,\th(y-x)W^{-1}(x)\bigg(\sbr{\int_{0}^{1}d\l^\prime \,{\rm e}^{i\,e\,\b_1\,\l^\prime\,\wti{F}(x)} T_a {\rm e}^{-i\,e\,\b_1\,\l^\prime\,\wti{F}(x)}}{K_1(\b_1;x)} \nonumber\\
&&+\int_{0}^{1}d\l^\prime D_1 \({\rm e}^{i\,e\,\b_1\,\l^\prime\,\wti{F}(x)}T_a{\rm e}^{-i\,e\,\b_1\,\l^\prime\,\wti{F}(x)}\)-ie\int_{0}^{1}d\l^\prime \,  {\rm e}^{i\,e\,\b_1\,\l^\prime\,\wti{F}(y)}\sbr{A_1}{T_a}{\rm e}^{-i\,e\,\b_1\,\l^\prime\,\wti{F}(y)}\nonumber\\
&&-i\,e\,\b_1\int_{0}^{1}d\l^\prime\int_{0}^{\l^\prime}d\l\sbr{ {\rm e}^{i\,e\,\b_1\,\l^\prime\,\wti{F}(x)}T_a{\rm e}^{-i\,e\,\b_1\,\l^\prime\,\wti{F}(x)}}{{\rm e}^{i\,e\,\b_1\,\l\,\wti{F}(x)} {\cal C}(x){\rm e}^{-i\,e\,\b_1\,\l\,\wti{F}(x)}}\bigg)W(x)\otimes\nonumber\\
&&\otimes \sbr{W^{-1}(y)K_1(\b_2;y)W(y)}{W^{-1}(x)T_a W(x)}\nonumber
\er
where $D_1 = \partial_1 + i\,e\,\sbr{A_1}{\cdot}$.

Collecting the expressions \rf{k1k2rel2} and \rf{irle4}, one can express \rf{qqpb} using \rf{aapb} in terms of explicit integrals, as follows
\be
\pbro{Q(\b_1)}{Q(\b_2)}=Q(\b_1)\otimes Q(\b_2)\({\cal Y} + \wti{\cal I}_L+\wti{\cal I}_R\)
\lab{qqpb2}
\ee
where we defined that
\br
\wti{\cal I}_{R/L} \equiv \int_{-\infty}^{+\infty} dx \int_{-\infty}^{+\infty} dy \,Q_1^{-1}(x)\otimes Q_2^{-1}(y)\,{\cal I}_{R/L}\,Q_1(x) \otimes Q_2(y)
\lab{tildei}
\er
and 
\br
{\cal Y} &\equiv& \int_{-\infty}^{+\infty} dx \int_{-\infty}^{+\infty} dy\,Q^{-1}_1(x)W^{-1}(x)\otimes Q^{-1}_2(y)W^{-1}(y)\times \nonumber\\
&&\times \pbro{K_1(\b_1,x)}{K_1(\b_2,y)}W(x) Q_1(x)\otimes W(y) Q(y).
\lab{y}
\er
Evaluating \rf{tildei} and \rf{y} (see \rf{yap}, \rf{ilap} and \rf{irap}), and substituting it into \rf{qqpb2} one finds that
\br
&&\pbro{Q(\b_1)}{Q(\b_2)} = \frac{e^2 }{\b_1 - \b_2}\bigg[\b_2^2 \(T_a -  V(\b_1)T_a V^{-1}(\b_1)\) Q(\b_1)\otimes \wti{\zeta}_a(\b_2,+\infty) Q(\b_2) \nonumber\\
&&+\b_1^2 \wti{\zeta}_a(\b_1,+\infty) Q(\b_1)\otimes \(T_a -  V(\b_2)T_aV^{-1}(\b_2)\) Q(\b_2)\nonumber\\
&&-\b_2^2\,Q(\b_1)Q^{-1}_R(\b_1)\(T_a -  V_R(\b_1)T_aV_R(\b_1)\) Q_R(\b_1)\otimes Q(\b_2)Q^{-1}_R(\b_2)\wti{\zeta}_a(\b_2,-\infty) Q_R(\b_2)\nonumber\\
&& - \b_1^2 \,Q(\b_1)Q^{-1}_R(\b_1)\wti{\zeta}_a(\b_1,-\infty) Q_R(\b_1)\otimes Q(\b_2)Q^{-1}_R(\b_2)\(T_a -  V_R(\b_2)T_aV_R^{-1}(\b_2)\) Q_R(\b_2)\bigg]\nonumber\\
&&-e^2\,\b_1 Q(\b_1)Q^{-1}_R(\b_1)\wti{\zeta}_a(\b_1,-\infty) Q_R(\b_1)\otimes\sbr{T_a}{Q(\b_2)Q^{-1}_R(\b_2)}  Q_R(\b_2)\nonumber\\
&&+e^2\,\b_2 \sbr{T_a}{Q(\b_1)Q^{-1}_R(\b_1)}Q_R(\b_1)\otimes Q^{-1}_R(\b_2)\wti{\zeta}_a(\b_2,-\infty) Q_R(\b_2)\nonumber\\
&&+Q(\b_1)\otimes Q(\b_2)\,{\cal Z}[{\cal C}]
\er
where  $V(\b) = V(\b,+\infty)= {\rm e}^{ie\b\wti{F}(+\infty)}$,  $V_R(\b) =V(\b,-\infty)  ={\rm e}^{ie\b\wti{F}(-\infty)}$,
${\cal Z}[{\cal C}]$ is expressed in \rf{zcal}, and
\be
\tilde{\zeta}_a(\b,x) \equiv \int_{0}^{1}d\l\,{\rm e}^{i\,e\,\b\,\l\,\wti{F}^W(x)} T_a {\rm e}^{-i\,e\,\b\,\l\,\wti{F}^W(x)}
\ee

Note that the following yields
\be
\b_1\wti{\zeta}_a(\b_1,-\infty) Q_R(\b_1)\otimes \(T_a -  V_R(\b_2)T_a V_R^{-1}(\b_2)\) = -\b_2 \(T_a - V_R(\b_1)T_a V_R(\b_1)\)\otimes \zeta_a(\b_2,-\infty)
\ee
which leads to
\br
&&\pbro{Q(\b_1)}{Q(\b_2)} =e^2\b_1 \wti{\zeta}_a(\b_1,+\infty) Q(\b_1)\otimes \(T_a -  V(\b_2)T_aV^{-1}(\b_2)\) Q(\b_2)\nonumber\\
&& - e^2\b_1\,Q(\b_1)Q^{-1}_R(\b_1)\wti{\zeta}_a(\b_1,-\infty) Q_R(\b_1)\otimes Q(\b_2)Q^{-1}_R(\b_2)\(T_a -  V_R(\b_2)T_aV_R^{-1}(\b_2)\) Q_R(\b_2)\nonumber\\
&&-e^2\,\b_1 Q(\b_1)Q^{-1}_R(\b_1)\wti{\zeta}_a(\b_1,-\infty) Q_R(\b_1)\otimes\sbr{T_a}{Q(\b_2)Q^{-1}_R(\b_2)}  Q_R(\b_2)\nonumber\\
&&+e^2\,\b_2 \sbr{T_a}{Q(\b_1)Q^{-1}_R(\b_1)}Q_R(\b_1)\otimes Q^{-1}_R(\b_2)\wti{\zeta}_a(\b_2,-\infty) Q_R(\b_2)\nonumber\\
&&+Q(\b_1)\otimes Q(\b_2)\,{\cal Z}[{\cal C}]
\er
When the constraints hold, that is, ${\cal C}=0$, the charge operators can be written as $Q(\b,x) = V(\b,x)$ and ${\cal Z} = 0$, hence one gets that
\br
&&\pbro{Q(\b_1)}{Q(\b_2)} \approx e^2\b_1 \wti{\zeta}_a(\b_1,+\infty) Q(\b_1)\otimes \sbr{T_a}{ Q(\b_2)}\nonumber\\
&&-e^2\,\b_1 Q(\b_1)Q^{-1}_R(\b_1)\wti{\zeta}_a(\b_1,-\infty) Q_R(\b_1)\otimes\sbr{T_a}{Q(\b_2)}  
\nonumber\\
&&+e^2\,\b_2 \sbr{T_a}{Q(\b_1)Q^{-1}_R(\b_1)}Q_R(\b_1)\otimes Q^{-1}_R(\b_2)\wti{\zeta}_a(\b_2,-\infty) Q_R(\b_2)
\er
Computing the transformation of the charge operator $Q(\b_1)$, under the charges $q_N(\b_2)$, one finds, under the imposition of the constraints, that
\be
\d_N Q(\b_1) \approx \vareps e^2\b_2 \sbr{T_a}{Q(\b_1)V^{-1}_R(\b_1)}V_R(\b_1)\,{\rm Tr}\(Q^N(\b_2) Q^{-1}_R(\b_2)\wti{\zeta}_a(\b_2,-\infty) Q_R(\b_2)\)
\lab{qqn}
\ee
where we used that
\be
\wti{\zeta}_a(\b_1,+\infty) Q(\b_1)\,{\rm}\(Q^N(\b_2) \sbr{T_a}{ Q(\b_2)}\) = 0
\ee
obtained from the adjoint action and from the fact that $\sbr{Q(\b_1)}{Q(\b_2)} =0$.

Comparing the r.h.s. of \rf{qqn} with \rf{xiref}, one finds that the charge operator has a global phase transformation under action of the conserved charges via Poisson brackets, given by
\be
\d_N Q(\b_1) \approx  \vareps e^2 \b_2\xi_a(N,\b_2;x_R)\,\sbr{T_a}{Q(\b_1)V_R^{-1}(\b_1)}V_R(\b_1).
\ee
where 
\be
\xi_a(N,\b,x) = {\rm Tr}\(Q^N(\b) Q^{-1}(\b,x)\wti{\zeta}_a(\b,x) Q(\b,x)\).
\ee

Furthermore, the Poisson algebra of the conserved charges $q(\b_1)$ and $q(\b_2)$, can be obtained from the result \rf{qqn} and considering \rf{sec2:cc}, such that we get the following symmetrized result: 
\be
\pbr{q_N(\b_1)}{q_M(\b_2)} \approx\vareps e^2 \b_2\xi_a(N,\b_2;x_R)\,{\rm Tr}\(Q^{M-1}(\b_2)\,\sbr{T_a}{Q(\b_1)V_R^{-1}(\b_1)}V_R(\b_1)\)
\ee

The conserved charges only Poisson commute, that is, they are in involution if the constant element $V_R$ is in the center of the gauge group, i.e., if $V_R \in Z(G)$, we have that
\be
\pbr{q_N(\b_1)}{q_M(\b_2)} \approx 0
\ee
Such a condition on $V_R(\b)$, due to the dependence on $\b$-parameter, leads $W_R^{-1}\wti{F}(-\infty)W_R$ to be an element of the center of the Lie-Algebra. Hence, the phase factor \rf{xiph} becomes functionally dependent only upon the charge operator:
\be
\xi(N,\b; x_R) \approx e^2\,\vareps\,\b {\rm Tr}\(Q^N(\b) T_a\).
\ee

Moreover, the Yang-Mills in $(1+1)$ dimensions does not share the same integrability-like structures as seen in the higher-dimensional settings \cite{ymintg}. There is no local Fundamental Poisson Lie bracket (FPR) in a loop space type-setting relation, since the connection ${\cal A}$ associated with the charge operator is non-local in the $0$-loop space, and consequently, there is no Sklyanin-type relation, however, the charge can be found in involution under appropriate boundary conditions and under the imposition of the constraints.

\section{Hidden symmetries of the integral equations}
\label{sec:6}
\setcounter{equation}{0}
In integrable models, the symmetries that preserve the zero curvature representation are fundamental in constructing new solutions of the dynamical equations. Such symmetries are neither symmetries of the dynamical equations of the system nor of the action, and so are called hidden symmetries. In fact, they correspond to a \emph{gauge-type} transformation defined by loop groups.

Regarding the two-dimensional Yang-Mills theory, the zero curvature representation is translated to a path-invariance condition on the eigenvalues of a line operator $U_{x_R}(\G)$ given in \rf{opdef}, guaranteed by the equation \rf{sec2:hinteq}. In fact, the functional of the eigenvalues ${\rm Tr}\(U^N(\G)\)$ should be invariant under the deformations $x \rightarrow x(\s) + \d x(\s)$ on the path $\G$ where $U(\G)$ is integrated, parameterized by $\s \in [\s_i, \s_f]$ with its endpoints $x_R=x(\s_i)$ and $x_f=x(\s_f)$ kept fixed, i.e., $\d x_R = \d x_f = 0$. One can verify such a condition by considering
\br
\d {\rm Tr}\(U^N(\G)\) = {\rm Tr}\(U^{N}U^{-1}\d U(\G)\).
\lab{defu}
\er
Note that we have omitted the dependence of $U$ upon the parameter $\b$ and the reference point $x_R$; still, the eigenvalues of the operator $U$ are independent of a reference point, and so is the trace. We have that, when we change the path with its endpoints kept fixed, the operator $U$ changes by (see Appendix \ref{apU})
\br
U^{-1} \d U(\s_f) &=&-\(U^{-1}W^{-1}\d W U\)(\G)
 + \int_{\s_i}^{\s_f}d\s\,U^{-1}(\s)\,W_\G^{-1}(\s)\,F_{\m \n}\,W_\G(\s)U(\s)\,\frac{dx^\m}{d\s}\d x^\n
\nonumber\\  
&+&\int_{\s_i}^{\s_f}d\s\,U^{-1}(\s) W^{-1}\(D_\n K_\m - D_\m K_\n - \sbr{K_\m}{K_\n}\)W U(\s)\,\frac{dx^\m}{d\s}\d x^\n.
\lab{duv}
\er
Using the expression \rf{sec2:lj} for $K_\m$ one can see that
\br
D_\n K_\m - D_\m K_\n  &=& ie^2\b \int_{0}^1d\l \,{\rm e}^{i\,e\,\l\,\b\,\wti{F}(x)}\(D_\m \wti{J}_\n - D_\n \wti{J}_\m\)\,{\rm e}^{-i\,e\,\l\,\b\,\wti{F}(x)}\nonumber\\
&+&ie^2\b \int_{0}^1d\l\bigg(\sbr{\(D_\m {\rm e}^{ie\l\b\wti{F}(x)}\){\rm e}^{-ie\l\b\wti{F}(x)} }{{\rm e}^{ie\l\b\wti{F}(x)}\wti{J}_\n{\rm e}^{-ie\l\b\wti{F}(x)}} \nonumber\\
&+&\sbr{{\rm e}^{ie\l\b\wti{F}(x)}\wti{J}_\m{\rm e}^{-ie\l\b\wti{F}(x)}}{\(D_\n {\rm e}^{ie\l\b\wti{F}(x)}\){\rm e}^{-ie\l\b\wti{F}(x)} } \bigg)
\lab{dkdk}
\er
with
\be
\(D_\m {\rm e}^{ie\l\b\wti{F}(x)}\){\rm e}^{-ie\l\b\wti{F}(x)} = ie\b \int_{0}^{\l}d\l^\prime \,{\rm e}^{ie\l^\prime\b\wti{F}(x)}D_\m \wti{F} {\rm e}^{-ie\l^\prime\b\wti{F}(x)}. 
\ee
Notice that, considering the relation
\be
\sbr{D_\m}{D_\n}\wti{F} = D_\m (D_\n \wti{F}) - D_\n (D_\m \wti{F})=ie\sbr{F_{\m \n}}{\wti{F}} = 0
\ee
and imposing the equations of motion \rf{sec2:em}, we obtain the covariant condition for the matter currents:
\be
D_\m \wti{J}_\n - D_\n \wti{J}_\m  = 0.
\lab{contjeq}
\ee
We can also write the following
\be
\(D_\m {\rm e}^{ie\l\b\wti{F}(x)}\){\rm e}^{-ie\l\b\wti{F}(x)} = ie^2\b \int_{0}^{\l}d\l^\prime \,{\rm e}^{ie\l^\prime\b\wti{F}(x)}\wti{J}_\m {\rm e}^{-ie\l^\prime\b\wti{F}(x)}.
\lab{dej}
\ee
Substituting both expressions \rf{contjeq} and \rf{dej} into \rf{dkdk}, we conclude that, when the equations of motion \rf{sec2:em} hold true, we obtain from the expression \rf{dkdk} that
\be
D_\m K_\n - D_\n K_\m - \sbr{K_\m}{K_\n} =0
\lab{nullcurv}
\ee
In addition, the operator $U(\s)$ commutes with $W_\G^{-1}(\s)\,F_{\m \n}\,W_\G(\s)$ when the integral dynamics equations \rf{sec2:hintaux} holds, then we have that 
\be
 \int_{\s_i}^{\s_f}d\s\,U^{-1}(\s)\,W_\G^{-1}(\s)\,F_{\m \n}\,W_\G(\s)U(\s)\,\frac{dx^\m}{d\s}\d x^\n = W^{-1}(\G)\d W(\G).
 \lab{comuw}
\ee
Hence, from the results \rf{nullcurv} and \rf{comuw}, the expression \rf{duv} becomes
\be
U^{-1}\d U(\G) = U^{-1}(\G)\sbr{U(\G)}{W^{-1}\d W(\G)}.
\lab{udu}
\ee
which correspond to the infinitesimal version of the transformation \rf{vH}, replacing $V$ by $U$, and where the element $W^{-1}\d W(\G)$ is associated with $H(\S)$.
Then, from \rf{udu}, the function of the eigenvalues of $U$, given by the trace ${\rm Tr}\(U^N\)$, is invariant under path-deformations, since
\be
\d{\rm Tr}\(U^N(\G)\) = {\rm Tr}\(\sbr{U^N(\G)}{W^{-1}\d W(\G)}\) = 0.
\lab{trun}
\ee
Furthermore, in $(3+1)$-dimensional Yang–Mills theory \cite{ymintg}, not only the eigenvalues of the operators associated with the conserved charges, but also the path-ordered operators themselves, satisfy path invariance. In contrast, in the two-dimensional model, one relies only on the path invariance of functions of the eigenvalues of the operators.

The equation \rf{nullcurv} holds for any point in spacetime, and so, along any point $x(\s)$ within the path $\G$ where the Wilson line is integrated. Hence, we can take an infinitesimal section in $\g \subset \G$, where the derivatives of $W(\g)$ are well defined, i.e.,
\be
\partial_\m W(\g) = -ie A_\m W(\g).
\ee
With such a condition, we can express the equation \rf{nullcurv} as a zero curvature equation:
\be
W^{-1}(\g)\(D_\n K_\m - D_\m K_\n - \sbr{K_\m}{K_\n}\)W(\g) = \partial_\m a_\n - \partial_\n a_\m - \sbr{a_\m}{a_\n} =0
\lab{nullcurv1}
\ee
where we defined that
\be
a_\m \equiv W^{-1} K_\m W.
\ee
which is in fact, the connection associated with the holonomy $U$, that is $a_\m \frac{dx^\m}{d\s}= \frac{dU}{d\s}U^{-1}$. Let $g$ be an element of the complexified gauge group $G^{\mathbb{C}}$, then any transformation of the kind (up to a constant element)
\be
U(\G) \rightarrow g(\G) \,U(\G) 
\lab{ugtr}
\ee
preserves the equation \rf{nullcurv1}, since it acts as gauge transformation on
the connection $a_\m(\s)$, i.e.,
\be
a_\m(\s) \rightarrow  g\,a_\m\,g^{-1} + \partial_\m g \, g^{-1}.
\ee
However, the condition of \rf{nullcurv1} is not sufficient to guarantee invariance under path deformations, since it is satisfied only when the trace of the deformed operator vanishes, as in \rf{trun}. Indeed, for the transformed operator obtained from \rf{ugtr}, one finds under path deformations that
\be
\d {\rm Tr}\big[ (g\,U)^N (\G)] = {\rm Tr} \big[\(g \,U\)^N U^{-1}\,g^{-1}\d g \,U + \(g \,U\)^{N}U^{-1}\sbr{U}{W^{-1} \d W(\G)}\big].
\lab{du}
\ee
Notice that the terms on the right-hand side of \rf{du} do not vanish separately. One must take into account that the variation $\d g$ depends on the original fields. Indeed, the expression \rf{du} vanishes only if this dependence is expressed in terms of the Wilson line $W(\G)$ associated with the original $A_\m$ field, and an arbitrary element $\wti{g} $, as follows
\be
g(\G) = W^{-1}(\G) \, \wti{g}(x_f)\, W(\G)
\lab{gwgt}
\ee
where we considered that $\wti{g}$ only depends upon the final point of $\G$, i.e., $\wti{g}(\G) = \wti{g}(x_f)$, such that, under path deformations with the endpoints of the path kept fixed, we should have $\d \wti{g}(x_f) = 0$. Since $g \in G^{\mathbb{C}}$ and $W \in G$, it follows that $\wti{g}$ is an element of the complexified Lie Group $G^{\mathbb{C}}$. Hence, the change of $g$ under path deformations is given by
\be
\d g = \sbr{g}{W^{-1}\d W},
\lab{dgaux}
\ee
Substituting \rf{dgaux} into \rf{du}, one finally finds that the function of the eigenvalues of the new operator is path-invariant, i.e.,
\be
\d {\rm Tr}\big[ (g\,U)^N (\G)] = 0.
\ee

Therefore, the transformations \rf{ugtr} constitute a hidden symmetry of the system, as they maintain the path-invariance condition. These symmetries act via the local transformations of the complexified gauge group $G^{\mathbb{C}}$ dressed by the Wilson lines of the original fields, see \rf{gwgt}. Moreover, such transformations may play a role analogous to the Kac-Moody group in integrable models in (1+1)-dimensions.

\section{Concluding remarks }
\label{sec:7}
\setcounter{equation}{0}
We have constructed an integral formulation of the local dynamical equations for Yang-Mills theories in a $(1+1)$-dimensional Minkowski spacetime, inspired by earlier developments in higher-dimensional settings \cite{lafgl1, lafgl2}. This formulation leads to the existence of gauge-invariant and dynamically conserved quantities, arising from the requirement of path-independence of the eigenvalues of the charge operator. In fact, we find that in this setting, there is no need to require the path-condition of holonomy operators, only on their eigenvalues.

Following the reasoning developed in \cite{ymintg}, we investigated whether these charges generate a novel symmetry of the theory by analyzing their action on phase space within the Hamiltonian formalism. We demonstrated that the charges are conserved in time and commute with the total Hamiltonian up to first-class constraints. As such, they generate transformations that preserve the physical dynamics of the system. Notably, these transformations act nontrivially on the canonical fields of the theory, identifying them as primary fields under this symmetry.

The charges constructed are both gauge-invariant and physical, as they commute with the constraints and are conserved under time evolution, in accordance with the criteria established in higher dimensions \cite{lafgl1,lafgl2}. Nevertheless, the structure in $(1+1)$ dimensions proves to be significantly simpler: the set of conserved charges does not involve high-dimensional integrations, and the requirement for path-independence is satisfied only for the eigenvalues of the holonomies present in the integral equations.

These results offer an interesting perspective for further investigation in the quantum regime. In particular, the simplified structure of the charges in two dimensions makes them especially suitable for study in the context of lattice gauge theories -- in fact, it can provide insights into the nature of the charges, since the non-local properties in the higher-dimensional settings present difficulties for the investigation in the quantum regime and require further analysis. As discussed in \cite{twolattice}, the behavior of these charges in the strong-coupling limit of two-dimensional QCD reveals a remarkable feature: color-singlet composite states such as mesons and baryons carry non-vanishing conserved charges, while isolated quarks do not. This supports the interpretation that the conserved charges are carried by physical, gauge-invariant states and may remain unconfined.

Although the two-dimensional setting is useful for gaining intuition about the nature of conserved charges due to its simplicity, it is not well-suited for investigating the structures found in higher-dimensional Yang-Mills theories that resemble integrability \cite{ymintg}, since such structures are absent in two dimensions.

\setcounter{equation}{0}

\vspace{2 cm}

\noindent {\bf Acknowledgements} We are grateful to Paulo A. Faria da Veiga and Ravi Mistry for many helpful discussions. LAF acknowledges the financial support of Fapesp
(Funda\c c\~ao de Amparo \`a Pesquisa do Estado de S\~ao Paulo) grant 2025/09036-5, and CNPq
(Conselho Nacional de Desenvolvimento Cient\'ifico e Tecnol\'ogico) grant 307833/2022-4. HM acknowledges the financial support of Fapesp grant 2021/10141-7.

\newpage
\appendix

\section{Evaluating the change on the operator $U$ induced by path changes}
\label{apU}
To evaluate $U^{-1} \d U$, consider the equation \rf{holeqq}, and that the deformation $\d$ is independent of the parameter $\s$, such that we obtain the following result
\be
\frac{d\d U}{d\s}- \d\(W_{\G}^{-1}\,K_\m\,W_\G \frac{dx^\m}{d\s}\) U\,\frac{dx^\m}{d\s} -W_{\G}^{-1}K_\m\,W_\G \, \d U\,\frac{dx^\m}{d\s} = 0
\lab{holeqqvar}
\ee
where, for the sake of notation, we omitted the dependence of $U$ upon $\b$ and $\s$. Admitting that $U_{x_R}$ has an inverse element that satisfies the following equation
\be
\frac{dU^{-1}}{d\s} + U^{-1}\(W_{\G}^{-1}(\s)\,K_\m(\b,x(\s))\,W_\G(\s) \, \frac{dx^\m}{d\s}\) = 0,
\lab{holeqqinv}
\ee
one can multiply \rf{holeqqvar} by $U^{-1}$ from the left, \rf{holeqqinv} by $\d U$ from the right, and summing them up, one gets that
\be
\frac{d}{d\s}\(U^{-1}\d U(\s)\) = U^{-1}(\s)\d\(W_{\G}^{-1}(\s)\,K_\m(x(\s))\,W_\G(\s) \frac{dx^\m}{d\s}\) U(\s).
\lab{dudu}
\ee
Since we are deforming $x(\s) \rightarrow x(\s)+ \d x (\s)$ keeping the endpoints of $\G$ fixed, i.e., $\d x(\s_i) = \d x(\s_f) = 0$, integrating \rf{dudu} on $\s \in [\s_i, \s_f]$, we get that
\be
U^{-1} \d U(\s_f) = \int_{\s_i}^{\s_f}d\s\,U^{-1}(\s) \(\d{\cal A}_\m\frac{dx^\m}{d\s} - \frac{d{\cal A}_\n}{d\s}\d x^\n - \sbr{{\cal A}_\m}{{\cal A}_\n}\frac{dx^\m}{d\s}\d x^\n\)U(\s)
\lab{dudu1}
\ee
where we defined that 
\be
{\cal A}_\m(\b,\s) = W_{\G}^{-1}(\s)\,K_\m(\b,x(\s))\,W_\G(\s)
\ee
Using the Wilson line differential equation \rf{sec2:holeq}, one gets that
\be
\frac{d{\cal A_\m}}{d\s} = W^{-1}_\G D_\n K_\m W_\G \frac{dx^\n}{d\s}.
\lab{dsa}
\ee
Additionally, considering that the action of $\d$ on local operators in spacetime acts as a partial derivative, we have that
\be
\d {\cal A}_\m = W^{-1}_\G \partial_\n K_\m W_\G \d x^\n + \sbr{W^{-1}_\G  K_\m W_\G}{W^{-1}_\G \d W_\G}.
\lab{da}
\ee
By the same reasoning used to derive \rf{dudu1}, we have that
\be
W^{-1}_\G\d W_\G(\s) = -ie\,W_{\G}^{-1}A_\m W_\G \d x^\m(\s) + \int_{\s_i}^{\s}d\s^\prime\,W_\G^{-1}\,F_{\m \n}\,W_\G\,\frac{dx^\m}{d\s^\prime}\d x^\n 
\lab{dw}
\ee
then, substituting \rf{dw} into \rf{da}, and  considering \rf{dsa}, the expression \rf{dudu1} becomes
\br
U^{-1} \d U(\s_f) &=& \int_{\s_i}^{\s_f}d\s\,U^{-1}(\s) W^{-1}\(D_\n K_\m - D_\m K_\n - \sbr{K_\m}{K_\n}\)W U(\s)\,\frac{dx^\m}{d\s}\d x^\n \nonumber\\
&+&\int_{\s_i}^{\s_f}d\s\,U^{-1}(\s)\sbr{W^{-1}_\G  K_\m W_\G}{\int_{\s_i}^{\s}d\s^\prime\,W_\G^{-1}\,F_{\m \n}\,W_\G\,\frac{dx^\m}{d\s^\prime}\d x^\n }U(\s)
\lab{dudu2}
\er
Taking into consideration the equation \rf{holeqq}, one can rewrite \rf{dudu2} as follows
\br
U^{-1} \d U(\s_f) &=&-\(U^{-1}W^{-1}\d W U\)(\G)
 + \int_{\s_i}^{\s_f}d\s\,U^{-1}(\s)\,W_\G^{-1}(\s)\,F_{\m \n}\,W_\G(\s)U(\s)\,\frac{dx^\m}{d\s}\d x^\n
\nonumber\\  
&+&\int_{\s_i}^{\s_f}d\s\,U^{-1}(\s) W^{-1}\(D_\n K_\m - D_\m K_\n - \sbr{K_\m}{K_\n}\)W U(\s)\,\frac{dx^\m}{d\s}\d x^\n.
\lab{dudu3}
\er

\section{A general expression for Poisson brackets with ho\-lo\-no\-mies}
\label{appendixA}
Given a functional of the physical fields $X$ and some holonomy $w$,  a solution of an ordinary differential equation, such as
\be
\frac{dw}{d\s} + \mathfrak{a}(\s)\,w(\s) = 0
\lab{auxholeq}
\ee
where $\mathfrak{a}(\s)$ is a functional of the physical fields. In addition, let us assume that there is an inverse $w^{-1}$ of $w$ that satisfies the ordinary differential equation
\be
\frac{dw^{-1}}{d\s} - w^{-1}(\s) \mathfrak{a}(\s) = 0
\lab{auxholeqinv}
\ee
Applying the Poisson brackets of $X$ on the equation \rf{auxholeq} we get that
\be
\{X\,,\,\frac{dw(\sigma)}{d\sigma}\} + \{X\,,\,\mathfrak{a}(\s)\}w\,+\mathfrak{a}(\sigma)\{X\,,\,w(\sigma)\} = 0.
\ee
Since the derivative $\frac{d}{d\sigma}$ commutes with the brackets, and $X$ does not depend upon $\sigma$, one can write the following
\be
\frac{d}{d\sigma} \{X\,,\,w(\sigma)\}+ \{X\,,\,\mathfrak{a}(\s)\}w+\mathfrak{a}(\s)\{X\,,\,w(\sigma)\} = 0
\lab{pbwx}
\ee
By multiplying the equation \rf{pbwx} on the left by $w^{-1}$, multiplying the equations \rf{auxholeqinv} on the right by $\{X\,,\,W(\sigma)\}$ and summing them up, the following
\be
\frac{d\;}{d\sigma}\(w^{-1}(\sigma)\{X\,,\,w(\sigma)\}\) +w^{-1}(\sigma)\,\{X\,,\,\mathfrak{a}(\s)\}\,w(\sigma)= 0  
\ee
Then integrating the result from $\sigma_i$ up to $\sigma$, and using that $w(\sigma_i)=w(x_R)=\one$, one then obtains that
\be
w^{-1}(\sigma)\,\{X\,,\,w(\sigma)\} = -\int_{\sigma_i}^{\sigma}d\sigma^\prime\,w^{-1}(\sigma^\prime)\,\{X\,,\,\mathfrak{a}(\s^\prime)\}\,w(\sigma^\prime)
\lab{auxpbwx}
\ee

If $\mathfrak{a} = ieA_\m \frac{dx^\m}{d\s}=ieA_1(x)$, where we considered that $\frac{dx^\m}{d\s}= \d_{\m 1}$ and $\s = x\in (-\infty, +\infty)$. The equation \rf{auxholeq} becomes the Wilson line's equation \rf{sholeq}, and using \rf{auxpbwx}, we have that
\be
W^{-1}(x)\,\{X\,,\,W(x)\} = -i\,e\,\int_{-\infty}^{x}dy\,W^{-1}(y)\,\{X\,,\,A_1(y)\}\,W(y)
\lab{pbwxf} 
\ee

If $\mathfrak{a} = - W^{-1}(\s) K_\m W(\s)\frac{dx^\mu}{d\s}= -W^{-1}(x)K_1(x)W(x)$, where we considered that $\frac{dx^\m}{d\s}= \d_{\m 1}$ and $\s = x\in (-\infty, +\infty)$. The equation \rf{auxholeq} becomes \rf{sholeq}, and using \rf{auxpbwx}, we have that
\be
\pbr{X}{Q(\b)} =Q(\b)\,\int_{-\infty}^{+\infty} dx\, Q^{-1}(x,-\infty)\pbr{X}{{\cal A}(\b;x)}\,Q(x,-\infty)
\lab{qrelA}.
\ee

\section{Calculations of the Poisson Algebra of the charges}
\label{appendixC}
In order to evaluate the integral \rf{y}, let us work on the expression \rf{k1k2rel2}. Using \rf{k1e}, one may integrate by parts the integrals in $\lambda_{1,2}$ variables, and find that
\br
&&\pbro{K_1(\b_1;x)}{K_1(\b_2;y)}=\d(x-y)\frac{e^2\b_1\b_2}{\b_1 - \b_2}\bigg(\sbr{{\rm e}^{i\,e\,\b_1\,\wti{F}}T_a {\rm e}^{-i\,e\,\b_1\,\wti{F}}}{K_1(\b_1,x)}\otimes \zeta_a(\b_2,x) \nonumber\\
&&+ \zeta_a(\b_1,x) \otimes \sbr{{\rm e}^{i\,e\,\b_1\,\wti{F}}T_a {\rm e}^{-i\,e\,\b_1\,\wti{F}}}{K_1(\b_2,x)}\bigg)\nonumber\\
&&+\d(x-y)\frac{e^2\b_1\b_2}{\b_1 - \b_2}\times\nonumber\\
&&\times \bigg\{\int_{0}^{1} d\l_1 \sbr{\frac{d}{d\l_1}\( {\rm e}^{i\,e\,\b_1\,\l_1\,\wti{F}} T_a  {\rm e}^{-i\,e\,\b_1\,\l_1\,\wti{F}}\)}{ie^2\b_1\int_{0}^{\l_1}d\l^\prime  {\rm e}^{i\,e\,\b_1\,\l^\prime\,\wti{F}}J_0  {\rm e}^{-i\,e\,\b_1\,\l^\prime\,\wti{F}}}\otimes \zeta_a(\b_2,x) \nonumber\\
&&+ \zeta_a(\b_1,x)\otimes \int_{0}^{1}d\l_2 \sbr{\frac{d}{d\l_2}\({\rm e}^{i\,e\,\b_2\,\l_2\,\wti{F}} T_a  {\rm e}^{-i\,e\,\b_2\,\l_2\,\wti{F}}\)}{ie^2\b_2\int_{0}^{\l_2}d\l^\prime {\rm e}^{ie\b_2\l^\prime \wti{F}}J_0{\rm e}^{-ie\b_2\l^\prime \wti{F}}}\bigg\}\nonumber\\
\lab{k1k2relap1}
\er
where we defined that
\be
\zeta_a(\b,x) \equiv \int_{0}^{1}d\l\,{\rm e}^{i\,e\,\b\,\l\,\wti{F}(x)} T_a {\rm e}^{-i\,e\,\b\,\l\,\wti{F}(x)}.
\lab{zr}
\ee
Using \rf{sec4:constraint} to write $J_0$ in terms of the constraints ${\cal C}$, and considering the integral:
\be
i\,e\,\b\int_{0}^{\l}d\l^\prime {\rm e}^{ie\b\l^\prime \wti{F}}D_1 \wti{F}{\rm e}^{-ie\b\l^\prime \wti{F}} = \(D_1 {\rm e}^{ie\b\l \wti{F}}\) {\rm e}^{-ie\b\l \wti{F}}
\ee
where $D_1 = \partial_1 + i\,e\,\sbr{A_1}{\cdot}$. One can rewrite the last two lines of the right-hand side of \rf{k1k2relap1} as follows
\br
&&\pbro{K_1(\b_1;x)}{K_1(\b_2;y)}=\d(x-y)\frac{e^2\b_1\b_2}{\b_1 - \b_2}\bigg(\sbr{{\rm e}^{i\,e\,\b_1\,\wti{F}}T_a {\rm e}^{-i\,e\,\b_1\,\wti{F}}}{K_1(\b_1,x)}\otimes \zeta_a(\b_2,x) \nonumber\\
&&+ \zeta_a(\b_1,x) \otimes \sbr{{\rm e}^{i\,e\,\b_1\,\wti{F}}T_a {\rm e}^{-i\,e\,\b_1\,\wti{F}}}{K_1(\b_2,x)}\bigg)\nonumber\\
&&+\d(x-y)\frac{e^2\b_1\b_2}{\b_1 - \b_2}\times\nonumber\\
&&\times \bigg\{\int_{0}^{1} d\l_1 \sbr{\frac{d}{d\l_1}\( {\rm e}^{i\,e\,\b_1\,\l_1\,\wti{F}} T_a  {\rm e}^{-i\,e\,\b_1\,\l_1\,\wti{F}}\)}{\(D_1 {\rm e}^{ie\b_1\l_1 \wti{F}}\) {\rm e}^{-ie\b_1\l_1 \wti{F}} }\otimes \zeta_a(\b_2,x) \nonumber\\
&&+ \zeta_a(\b_1,x)\otimes \int_{0}^{1}d\l_2 \sbr{\frac{d}{d\l_2}\({\rm e}^{i\,e\,\b_2\,\l_2\,\wti{F}} T_a  {\rm e}^{-i\,e\,\b_2\,\l_2\,\wti{F}}\)}{\(D_1 {\rm e}^{ie\b_2\l_2 \wti{F}}\) {\rm e}^{-ie\b_2\l_2 \wti{F}}}\bigg\}\nonumber\\
&&-\d(x-y){\cal M}[{\cal C}]\nonumber\\
\lab{k1k2relap2}
\er
where we defined that
\br
&&{\cal M}[{\cal C}] \equiv \frac{e^2\b_1\b_2}{\b_1 - \b_2}\times\lab{calm}\\
&&\times \bigg\{\int_{0}^{1} d\l_1 \sbr{\frac{d}{d\l_1}\( {\rm e}^{i\,e\,\b_1\,\l_1\,\wti{F}} T_a  {\rm e}^{-i\,e\,\b_1\,\l_1\,\wti{F}}\)}{ie\b_1\int_{0}^{\l_1}d\l^\prime  {\rm e}^{i\,e\,\b_1\,\l^\prime\,\wti{F}}{\cal C}  {\rm e}^{-i\,e\,\b_1\,\l^\prime\,\wti{F}}}\otimes \zeta_a(\b_2,x) \nonumber\\
&&+ \zeta_a(\b_1,x)\otimes \int_{0}^{1}d\l_2 \sbr{\frac{d}{d\l_2}\({\rm e}^{i\,e\,\b_2\,\l_2\,\wti{F}} T_a  {\rm e}^{-i\,e\,\b_2\,\l_2\,\wti{F}}\)}{ie\b_2\int_{0}^{\l_2}d\l^\prime {\rm e}^{ie\b_2\l^\prime \wti{F}}{\cal C}{\rm e}^{-ie\b_2\l^\prime \wti{F}}}\bigg\}\nonumber
\er
Notice that
\be
\frac{d}{d\l}\({\rm e}^{ie\b\l \wti{F}}T_a{\rm e}^{-ie\b\l \wti{F}}\) = ie\b{\rm e}^{ie\b\l \wti{F}}\sbr{\wti{F}}{T_a}{\rm e}^{-ie\b\l \wti{F}}
\lab{incol1}
\ee
and
\be
\int_{0}^1 d\l \,\frac{d}{d\l}\({\rm e}^{ie\b\l \wti{F}}T_a{\rm e}^{-ie\b\l \wti{F}}\) = {\rm e}^{ie\b \wti{F}}T_a{\rm e}^{-ie\b \wti{F}} -T_a 
\lab{incol2}
\ee
In addition, using \rf{zr}, \rf{incol2}, and from the properties of the adjoint action, one finds that
\br
\sbr{\wti{F}}{T_a}\otimes \zeta_a(\b_2,x) &=& T_a\otimes \int_{0}^1d\l_2{\rm e}^{ie\b_2\l_2 \wti{F}}\sbr{T_a}{\wti{F}}{\rm e}^{-ie\b_2\l_2 \wti{F}}\nonumber\\
&=&\frac{i}{e\b_2}T_a\otimes \({\rm e}^{ie\b_2 \wti{F}}T_a{\rm e}^{-ie\b_2 \wti{F}} - T_a\)
\lab{propapb}
\er
Hence, using \rf{incol1}, \rf{incol2} and \rf{propapb}, one can rewrite \rf{k1k2relap2} as follows
\br
&&\pbro{K_1(\b_1;x)}{K_1(\b_2;y)}=\nonumber\\
&&=\d(x-y)\frac{e^2\b_1\b_2}{\b_1 - \b_2}\bigg(\sbr{T_a}{K_1(\b_1,x)}\otimes \zeta_a(\b_2,x) + \zeta_a(\b_1,x) \otimes \sbr{T_a}{K_1(\b_2,x)}\bigg)\nonumber\\
&&+\d(x-y)\frac{e^2\b_1\b_2}{\b_1 - \b_2}\bigg\{\frac{\b_1}{\b_2}\sbr{\zeta_{a}(\b_1,x)}{K_1(\b_1,x)}\otimes\(T_a - {\rm e}^{ie\b_2\wti{F}}T_a{\rm e}^{-ie\b_2\wti{F}}\)  \nonumber\\
&&+\frac{\b_2}{\b_1}\(T_a - {\rm e}^{ie\b_1\wti{F}}T_a{\rm e}^{-ie\b_1\wti{F}}\)\otimes \sbr{\zeta_a(\b_2,x)}{K_1(\b_2,x)}\nonumber\\
&&+\frac{\b_1}{\b_2}\(D_1\zeta_a(\b_1,x)- ie \int_{0}^1d\l_1 {\rm e}^{ie\b_1\l_1\wti{F}}\sbr{A_1}{T_a} {\rm e}^{-ie\b_1\l_1 \wti{F}}\)\otimes\(T_a - {\rm e}^{ie\b_2\wti{F}}T_a {\rm e}^{-ie\b_2 \wti{F}}\) \nonumber\\
&&+\frac{\b_2}{\b_1} \(T_a - {\rm e}^{ie\b_1\wti{F}}T_a {\rm e}^{-ie\b_1 \wti{F}}\)\otimes \(D_1\zeta_a(\b_2,x) - ie \int_{0}^1d\l_2 {\rm e}^{ie\b_2\l_2\wti{F}}\sbr{A_1}{T_a} {\rm e}^{-ie\b_2\l_2 \wti{F}}\)\bigg\}\nonumber\\
&&-\d(x-y){\cal M}[{\cal C}]\nonumber\\
\lab{k1k2relap3}
\er
where we used that
\be
\int_{0}^1d\l \sbr{{\rm e}^{ie\b \l \wti{F}}T_a{\rm e}^{-ie\b \l \wti{F}}}{(D_1 {\rm e}^{ie\b \l \wti{F}}){\rm e}^{-ie\b \l \wti{F}}} =D_1\zeta_a(\b,x)- ie \int_{0}^1d\l {\rm e}^{ie\b\l\wti{F}}\sbr{A_1}{T_a} {\rm e}^{-ie\b\l \wti{F}}
\ee
Now, defining the quantity
\br
\tilde{\zeta}_a(\b,x) &\equiv& \int_{0}^{1}d\l\,{\rm e}^{i\,e\,\b\,\l\,\wti{F}^W(x)} T_a {\rm e}^{-i\,e\,\b\,\l\,\wti{F}^W(x)}
\nonumber\\
&=& W^{-1}\(D_1\zeta_a(\b,x)- ie \int_{0}^1d\l {\rm e}^{ie\b\l\wti{F}}\sbr{A_1}{T_a} {\rm e}^{-ie\b\l \wti{F}}\)W 
\er
where we denoted that $\wti{F}^W = W^{-1}\wti{F}W$.
When substituting \rf{k1k2relap3} into \rf{y}, and using \rf{holequ} and \rf{sholeq}, one obtains that
\br
&&{\cal Y} = \frac{e^2 \b_1 \b_2}{\b_1 - \b_2}\int_{-\infty}^{+\infty}dx \,\bigg[\frac{d}{dx}\(Q_1^{-1}(x) T_a Q_1(x)\)\otimes Q_2^{-1}(x)\wti{\zeta}_a(\b_2,x)Q_2(x) \nonumber\\ 
&&+Q_1^{-1}(x)\wti{\zeta}_a(\b_1,x)Q_1(x)\otimes\frac{d}{dx}\(Q_2^{-1}(x)T_aQ_2(x)\)\bigg]\nonumber\\
&&+ \frac{e^2 }{\b_1 - \b_2}\int_{-\infty}^{+\infty}dx\bigg[\b_1^2\frac{d}{dx}\(Q^{-1}_1(x)\wti{\zeta}_a(\b_1,x)Q_1(x)\)\otimes Q^{-1}_2(x)\(T_a - {\rm e}^{ie\b_2\wti{F}^W}T_a {\rm e}^{-ie\b_2 \wti{F}^W}\)Q_2(x) \nonumber\\
&&+ \b_2^2Q_1^{-1}(x)\(T_a - {\rm e}^{ie\b_1\wti{F}^W}T_a {\rm e}^{-ie\b_1 \wti{F}^W}\)Q_1(x)\otimes\frac{d}{dx}\(Q^{-1}_2(x)\wti{\zeta}_a(\b_2,x)Q_2(x)\)\nonumber\\
&&-\int_{-\infty}^{+\infty}dx\, Q^{-1}_1(x) W^{-1}(x)\otimes Q^{-1}_2(x)W^{-1}(x)\,{\cal M}\,[{\cal C}]W(x)Q_1(x)\otimes W(x)Q_2(x)
\lab{yapp}
\er
where we denoted $Q_{s}(x) = Q(\b_s,x)$ for $s=1,2$.

Performing an integration by parts in \rf{yapp}, one finds that
\br
&&{\cal Y} = \frac{e^2 \b_1 \b_2}{\b_1 - \b_2}\bigg(Q^{-1}(\b_1)T_a Q(\b_1)\otimes Q^{-1}(\b_2)\wti{\zeta}_a(\b_2,+\infty) Q(\b_2) \nonumber\\
&&+Q^{-1}(\b_1)\wti{\zeta}_a(\b_1,+\infty) Q(\b_1)\otimes Q^{-1}(\b_2)T_a Q(\b_2)\nonumber\\
&&-Q^{-1}_R(\b_1)T_a Q_R(\b_1)\otimes Q^{-1}_R(\b_2)\wti{\zeta}_a(\b_2,-\infty) Q_R(\b_2)\nonumber\\
&& - Q^{-1}_R(\b_1)\wti{\zeta}_a(\b_1,-\infty) Q_R(\b_1)\otimes Q^{-1}_R(\b_2)T_a Q_R(\b_2)\bigg)\nonumber\\
&&+e^2\int_{-\infty}^{+\infty}dx\bigg[\b_1\frac{d}{dx}\(Q^{-1}_1(x)\wti{\zeta}_a(\b_1,x)Q_1(x)\)\otimes Q^{-1}_2(x)T_a Q_2(x) \nonumber\\
&&- \b_2Q_1^{-1}(x)T_a Q_1(x)\otimes\frac{d}{dx}\(Q^{-1}_2(x)\wti{\zeta}_a(\b_2,x)Q_2(x)\)\nonumber\\
&&- \frac{e^2 }{\b_1 - \b_2}\bigg\{\b_1^2\bigg(Q^{-1}(\b_1)\wti{\zeta}_a(\b_1,+\infty)Q(\b_1)\otimes Q^{-1}(\b_2) {\rm e}^{ie\b_2\wti{F}^W(+\infty)}T_a {\rm e}^{-ie\b_2 \wti{F}^W(+\infty)}Q(\b_2)\nonumber\\
&&- Q^{-1}_R(\b_1)\wti{\zeta}_a(\b_1,-\infty)Q_R(\b_1)\otimes Q^{-1}_R(\b_2) {\rm e}^{ie\b_2\wti{F}^W(x_R)}T_a {\rm e}^{-ie\b_2 \wti{F}^W(x_R)}Q_R(B_2) \bigg)\nonumber\\
&&+ \b_2^2\bigg(Q^{-1}(\b_1) {\rm e}^{ie\b_1\wti{F}^W(+\infty)}T_a {\rm e}^{-ie\b_1 \wti{F}^W(+\infty)}Q(\b_1)\otimes Q^{-1}(\b_2)\wti{\zeta}_a(\b_2,+\infty)Q(\b_2)\nonumber\\
&&-Q_R^{-1}(\b_1) {\rm e}^{ie\b_1\wti{F}^W(-\infty)}T_a {\rm e}^{-ie\b_1 \wti{F}^W(-\infty)}Q_R(\b_1)\otimes Q^{-1}_R(\b_2)\wti{\zeta}_a(\b_2,-\infty)Q_R(\b_2)\bigg)\bigg\}\nonumber\\
&&-\wti{\cal M}[{\cal C}]
\lab{yap}
\er
where we defined
\br
\wti{\cal M}[{\cal C}] &\equiv& \int_{-\infty}^{+\infty}dx\, Q^{-1}_1(x) W^{-1}(x)\otimes Q^{-1}_2(x)W^{-1}(x)\,{\cal M}\,[{\cal C}]W(x)Q_1(x)\otimes W(x)Q_2(x)\nonumber\\
&-& \frac{e^2 }{\b_1 - \b_2}\int_{-\infty}^{+\infty}dx\bigg[\b_1^2 \,Q^{-1}_1(x)\wti{\zeta}_a(\b_1,x)Q_1(x)\otimes \nonumber\\
&\otimes& Q^{-1}_2(x) {\rm e}^{ie\b_2\wti{F}^W}\sbr{W^{-1}\(L_1 - K_1\)W}{T_a} {\rm e}^{-ie\b_2 \wti{F}^W}Q_2(x) \nonumber\\
&+& \b_2^2Q_1^{-1}(x) {\rm e}^{ie\b_1\wti{F}^W}\sbr{W^{-1}\(L_1 - K_1\)W}{T_a} {\rm e}^{-ie\b_1 \wti{F}^W}Q_1(x)\otimes\nonumber\\
&\otimes&\frac{d}{dx}\(Q^{-1}_2(x)\wti{\zeta}_a(\b_2,x)Q_2(x)\)\bigg]
\er
The expression of $\wti{\cal I}_{R/L}$ can be given explicitly by substituting \rf{irle4} into \rf{tildei}, that is
\br
\wti{\cal I}_L&=&e^2\,\b_2\,\int_{-\infty}^{+\infty}dx\int_{-\infty}^{x}dy\,Q^{-1}_1(x)\otimes Q^{-1}_2(y)\times \nonumber\\
&\times&\bigg\{\sbr{W^{-1}(x)K_1(\b_1;x)W(x)}{T_a } \otimes \bigg( \frac{d\wti{\zeta}_a(\b_2,y)}{dy}+\sbr{\wti{\zeta}_a(\b_2,y)}{W^{-1}K_1(\b_2,y)W}\nonumber\\
&-&i\,e\,\b_2\,\int_{0}^{1}d\l^\prime\int_{0}^{\l^\prime}d\l\bigg[ {\rm e}^{i\,e\,\b_2\,\l^\prime\,\wti{F}^W(y)}T_a{\rm e}^{-i\,e\,\b_2\,\l^\prime\,\wti{F}^W(y)}\,,\nonumber\\
&&\,,\,{\rm e}^{i\,e\,\b_2\,\l\,\wti{F}^W(y)} W^{-1}{\cal C}(y)W{\rm e}^{-i\,e\,\b_2\,\l\,\wti{F}^W(y)}\bigg]\bigg) \bigg\}  Q_1(x)\otimes Q_2(y)\nonumber\\
\lab{irle5}\\
\wti{\cal I}_R&=&-e^2\,\b_1\int_{-\infty}^{+\infty}dy\int_{-\infty}^{y}dx\,Q^{-1}_1(x)\otimes Q^{-1}_2(y)\bigg\{\bigg(\frac{d \wti{\zeta}_a(\b_1,x)}{dx}+\sbr{\wti{\zeta}_a(\b_1,x)}{W^{-1}K_1(\b_1;x)W} \nonumber\\
&-&i\,e\,\b_1\int_{0}^{1}d\l^\prime\int_{0}^{\l^\prime}d\l\bigg[ {\rm e}^{i\,e\,\b_1\,\l^\prime\,\wti{F}^W(x)}T_a{\rm e}^{-i\,e\,\b_1\,\l^\prime\,\wti{F}^W(x)}\,\,\nonumber\\
&&\,,\,{\rm e}^{i\,e\,\b_1\,\l\,\wti{F}^W(x)}W^{-1} {\cal C}(x)W{\rm e}^{-i\,e\,\b_1\,\l\,\wti{F}^W(x)}\bigg]\bigg)\otimes\nonumber\\
&\otimes&\sbr{W^{-1}(y)K_1(\b_2;y)W(y)}{T_a}\bigg\} Q_1(x) \otimes Q_2(y)\nonumber
\er
where we denoted $Q_{s}(x) = Q(\b_s,x)$ for $s=1,2$, and $\wti{F}^W = W^{-1}\wti{F}W$. Notice that, from the holonomy equation of $Q(\b,x)$ i.e. \rf{holequ}, we can obtain the following relations:
\br
\frac{d}{dx}\(Q^{-1}(\b,x) T_a Q(\b,x)\) &=& Q^{-1}(\b,x)\sbr{T_a}{W^{-1} K_1(\b,x)W}Q(\b,x)\nonumber\\
\lab{relqtq}\\
\frac{d}{dx}\bigg[Q^{-1}(\b,x)\wti{\zeta}_a(\b,x)Q(\b,x)\bigg]&=&Q^{-1}(\b,x)\bigg(\frac{d\wti{\zeta}_a(\b,x)}{dx} +\nonumber\\
&+&\sbr{\wti{\zeta}_a(\b,x)}{W^{-1}K_1(\b;x)W}\bigg)Q(\b,x)\nonumber
\er
Using \rf{relqtq} into \rf{irle5} and integrating the expressions in the r.h.s. of $\wti{\cal I}_L$ that do not involve the constraints ${\cal C}$, we obtain that 
\br
&&\wti{\cal I}_L = -e^2\,\b_2\,\int_{-\infty}^{+\infty}dx \frac{d}{dx}\(Q^{-1}_1(x) T_a Q_1(x)\)\otimes \bigg(Q^{-1}_2(x)\wti{\zeta}_a(\b_2,x)Q_2(x) -\nonumber\\
&&-Q^{-1}_R(\b_2)\wti{\zeta}_a(\b_2,x_R)Q_R(\b_2)\bigg)\nonumber\\
&&+ie^3\,\b_2^2\int_{-\infty}^{+\infty}dx\, \frac{d}{dx}\(Q_1^{-1}(x) T_a Q_2(x)\)\otimes \nonumber\\
&&\otimes \int_{-\infty}^{x}dy\,Q^{-1}_2(y)\bigg(\int_{0}^{1}d\l^\prime\int_{0}^{\l^\prime}d\l\bigg[ {\rm e}^{i\,e\,\b_2\,\l^\prime\,\wti{F}^W(y)}T_a{\rm e}^{-i\,e\,\b_2\,\l^\prime\,\wti{F}^W(y)}\,,\nonumber\\
&&\,,\,{\rm e}^{i\,e\,\b_2\,\l\,\wti{F}^W(y)} W^{-1}{\cal C}(y)W{\rm e}^{-i\,e\,\b_2\,\l\,\wti{F}^W(y)}\bigg]\bigg)Q_2(y)
\er
where $Q(\b)$ is the holonomy at the reference point $x_R= (t, -\infty)$, i.e., $Q_R(\b) = Q(\b,x_R)$.
Performing the first integral in the r.h.s., we get that
\br
\wti{\cal I}_L &=& -e^2\,\b_2\,Q^{-1}(\b_1) T_a Q(\b_1)\otimes Q^{-1} (\b_2)\wti{\zeta}_a(\b_2,+\infty)Q(\b_2)\nonumber\\
&+&e^2\,\b_2\,Q^{-1}(\b_1) T_a Q(\b_1)\otimes Q^{-1}_R(\b_2)\wti{\zeta}_a(\b_2,x_R)Q_R(\b_2)\nonumber\\
&+&e^2\,\b_2 \int_{-\infty}^{+\infty}dx \,Q^{-1}_1(x) T_a Q_1(x)\otimes \frac{d}{dx}\(Q^{-1}_2(x)\wti{\zeta}_a(\b_2,x)Q_2(x)\)\nonumber\\
&+&i\,e^3\,\b_2^2\int_{-\infty}^{+\infty}dx\, \frac{d}{dx}\(Q_1^{-1}(x) T_a Q_2(x)\)\otimes \nonumber\\
&\otimes& \int_{-\infty}^{x}dy\,Q^{-1}_2(y)\bigg(\int_{0}^{1}d\l^\prime\int_{0}^{\l^\prime}d\l\bigg[ {\rm e}^{i\,e\,\b_2\,\l^\prime\,\wti{F}^W(y)}T_a{\rm e}^{-i\,e\,\b_2\,\l^\prime\,\wti{F}^W(y)}\,,\nonumber\\
&&\,,\,{\rm e}^{i\,e\,\b_2\,\l\,\wti{F}^W(y)} W^{-1}{\cal C}(y)W{\rm e}^{-i\,e\,\b_2\,\l\,\wti{F}^W(y)}\bigg]\bigg)Q_2(y)
\lab{ilap}
\er
where $Q(\b_1)$ corresponds to the holonomy integrated on the spatial direction up to $+\infty$, i.e., $Q(\b_1) = Q(\b_1,+\infty)$.
The same procedure done for $\wti{\cal I}_R$ in \rf{irle5}, gives the follow
\br
\wti{\cal I}_R &=& e^2\,\b_1\,Q^{-1}(\b_1)\wti{\zeta}_a(\b_1,+\infty)Q(\b_1)\otimes Q^{-1}(\b_2)T_a Q(\b_2)\nonumber\\
&-&e^2\,\b_1\,Q^{-1}_R(\b_1)\wti{\zeta}_a(\b_1,x_R)Q_R(\b_1)\otimes Q^{-1}(\b_2)T_a Q(\b_2)\nonumber\\
&-&e^2\,\b_1 \int_{-\infty}^{+\infty}dy\,\frac{d}{dy}\(Q^{-1}(\b_1)\wti{\zeta}_a(\b_1,x)Q(\b_1)\)\otimes Q_2^{-1}(y)T_a Q_2(y)\nonumber\\
&-&ie^3\,\b_1^2 \int_{-\infty}^{+\infty}dy\int_{-\infty}^{y}dx Q^{-1}_1(x)\bigg(\int_{0}^{1}d\l^\prime\int_{0}^{\l^\prime}d\l\bigg[ {\rm e}^{i\,e\,\b_1\,\l^\prime\,\wti{F}^W(x)}T_a{\rm e}^{-i\,e\,\b_1\,\l^\prime\,\wti{F}^W(x)}\,\,\nonumber\\
&&\,,\,{\rm e}^{i\,e\,\b_1\,\l\,\wti{F}^W(x)}W^{-1} {\cal C}(x)W{\rm e}^{-i\,e\,\b_1\,\l\,\wti{F}^W(x)}\bigg]\bigg)Q_1(x)\otimes\nonumber\\
&\otimes&\frac{d}{dy}\(Q^{-1}_2(y) T_a Q_2(y)\).
\lab{irap}
\er
where $Q(\b_2)$ corresponds to the holonomy integrated on the spatial direction up to $+\infty$, i.e., $Q(\b_2) = Q(\b_2,+\infty)$.

Therefore, the sum of \rf{yap}, \rf{ilap}, and \rf{irap} results
\br
&&{\cal Y}+ \wti{\cal I}_L + \wti{\cal I}_R = \frac{e^2 }{\b_1 - \b_2}\times \nonumber\\
&&\times\bigg[\b_2^2 Q^{-1}(\b_1)\(T_a -  {\rm e}^{ie\b_1\wti{F}^W(+\infty)}T_a {\rm e}^{-ie\b_1 \wti{F}^W(+\infty)}\) Q(\b_1)\otimes Q^{-1}(\b_2)\wti{\zeta}_a(\b_2,+\infty) Q(\b_2) \nonumber\\
&&+\b_1^2 Q^{-1}(\b_1)\wti{\zeta}_a(\b_1,+\infty) Q(\b_1)\otimes Q^{-1}(\b_2)\(T_a -  {\rm e}^{ie\b_2\wti{F}^W(+\infty)}T_a {\rm e}^{-ie\b_2 \wti{F}^W(+\infty)}\) Q(\b_2)\nonumber\\
&&-\b_2^2Q^{-1}_R(\b_1)\(T_a -  {\rm e}^{ie\b_1\wti{F}^W(-\infty)}T_a {\rm e}^{-ie\b_1 \wti{F}^W(-\infty)}\) Q_R(\b_1)\otimes Q^{-1}_R(\b_2)\wti{\zeta}_a(\b_2,-\infty) Q_R(\b_2)\nonumber\\
&& - \b_1^2 Q^{-1}_R(\b_1)\wti{\zeta}_a(\b_1,-\infty) Q_R(\b_1)\otimes Q^{-1}_R(\b_2)\(T_a -  {\rm e}^{ie\b_2\wti{F}^W(-\infty)}T_a {\rm e}^{-ie\b_2 \wti{F}^W(-\infty)}\) Q_R(\b_2)\bigg]\nonumber\\
&&-e^2\,\b_1Q^{-1}_R(\b_1)\wti{\zeta}_a(\b_1,-\infty) Q_R(\b_1)\otimes\( Q^{-1}(\b_2)T_a  Q(\b_2)-Q^{-1}_R(\b_2)T_a  Q_R(\b_2)\ \)\nonumber\\
&&+e^2\,\b_2 \(Q^{-1}(\b_1) T_a Q(\b_1) - Q^{-1}_R(\b_1)T_aQ_R(\b_1)\)\otimes Q^{-1}_R(\b_2)\wti{\zeta}_a(\b_2,-\infty) Q_R(\b_2)\nonumber\\
&&+{\cal Z}[{\cal C}]
\lab{sum}
\er
where we defined
\br
{\cal Z}[{\cal C}]&=& \wti{M}[{\cal C}]+ i\,e^3\,\b_2^2\int_{-\infty}^{+\infty}dx\, \frac{d}{dx}\(Q_1^{-1}(x) T_a Q_2(x)\)\otimes \nonumber\\
&\otimes& \int_{-\infty}^{x}dy\,Q^{-1}_2(y)\bigg(\int_{0}^{1}d\l^\prime\int_{0}^{\l^\prime}d\l\bigg[ {\rm e}^{i\,e\,\b_2\,\l^\prime\,\wti{F}^W(y)}T_a{\rm e}^{-i\,e\,\b_2\,\l^\prime\,\wti{F}^W(y)}\,,\nonumber\\
&&\,,\,{\rm e}^{i\,e\,\b_2\,\l\,\wti{F}^W(y)} W^{-1}{\cal C}(y)W{\rm e}^{-i\,e\,\b_2\,\l\,\wti{F}^W(y)}\bigg]\bigg)Q_2(y)\nonumber\\
&-&ie^3\,\b_1^2 \int_{-\infty}^{+\infty}dy\int_{-\infty}^{y}dx Q^{-1}_1(x)\bigg(\int_{0}^{1}d\l^\prime\int_{0}^{\l^\prime}d\l\bigg[ {\rm e}^{i\,e\,\b_1\,\l^\prime\,\wti{F}^W(x)}T_a{\rm e}^{-i\,e\,\b_1\,\l^\prime\,\wti{F}^W(x)}\,\,\nonumber\\
&&\,,\,{\rm e}^{i\,e\,\b_1\,\l\,\wti{F}^W(x)}W^{-1} {\cal C}(x)W{\rm e}^{-i\,e\,\b_1\,\l\,\wti{F}^W(x)}\bigg]\bigg)Q_1(x)\otimes\frac{d}{dy}\(Q^{-1}_2(y) T_a Q_2(y)\).
\lab{zcal}
\er

\end{document}